\documentclass[journal=jacsat,manuscript=article]{achemso}
\usepackage{chemformula} % Formula subscripts using \ch{}
\usepackage[T1]{fontenc} % Use modern font encodings

\usepackage{longtable} % tables with pagebreak
\usepackage{multirow} % combine cells in table
\usepackage{amssymb} % additional mathematic symbols
\usepackage{braket} % braket notation
\usepackage{amsmath} % multi-line equation 
\usepackage{float}
\usepackage[figuresright]{rotating} % rotating figures
\usepackage{hyperref} % create clickable hyperlinks

\usepackage{verbatim} %block comment

\usepackage{soul} % highlight text using \hl{}

\usepackage{xcolor} % text color
\usepackage{multicol} %  make stuff single/multi-column as needed
\let\oldmaketitle\maketitle
\let\maketitle\relax

\SectionNumbersOn

\newcommand{\nocontentsline}[3]{}
\newcommand{\tocless}[2]{\bgroup\let\addcontentsline=\nocontentsline#1{#2}\egroup}

\author{Tsai-Jung Liu}
    \altaffiliation{Equal contribution}
    \affiliation{Faculty of Chemistry and Food Chemistry, TU Dresden, Bergstraße 66c, 01069 Dresden, Germany}
\author{Florian M. Arnold}
    \altaffiliation{Equal contribution}
    \affiliation{Faculty of Chemistry and Food Chemistry, TU Dresden, Bergstraße 66c, 01069 Dresden, Germany}
\author{Alireza Ghasemifard}
    \affiliation{Faculty of Chemistry and Food Chemistry, TU Dresden, Bergstraße 66c, 01069 Dresden, Germany}
\author{Qing-Long Liu}
    \affiliation{Faculty of Chemistry and Food Chemistry, TU Dresden, Bergstraße 66c, 01069 Dresden, Germany}
\author{Dorothea Golze}
    \affiliation{Faculty of Chemistry and Food Chemistry, TU Dresden, Bergstraße 66c, 01069 Dresden, Germany} 
\author{Agnieszka Kuc}
    \affiliation{Helmholtz-Zentrum Dresden-Rossendorf, Bautzner Landstr. 400, 01328 Dresden, Germany}
    \alsoaffiliation{Center for Advanced Systems Understanding, Conrad-Schiedt-Straße 20, 02826 G\"orlitz, Germany}
\author{Thomas Heine}
    \email{thomas.heine@tu-dresden.de}
    \affiliation{Faculty of Chemistry and Food Chemistry, TU Dresden, Bergstraße 66c, 01069 Dresden, Germany}
    \alsoaffiliation{Helmholtz-Zentrum Dresden-Rossendorf, Bautzner Landstr. 400, 01328 Dresden, Germany}
    \alsoaffiliation{Center for Advanced Systems Understanding, Conrad-Schiedt-Straße 20, 02826 G\"orlitz, Germany}
    \alsoaffiliation{Department of Chemistry and IBS Center for NanoMedicine, Yonsei University, Seodaemun-gu, Seoul 120-749, Republic of Korea}

\title{Electronic Structure and Topology in Gulf-edged Zigzag Graphene Nanoribbons}

\begin{document}

%change later or delete
%\begin{tocentry}
%Some journals require a graphical entry for the Table of Contents.
%This should be laid out ``print ready'' so that the sizing of the text is correct.
%Inside the \texttt{tocentry} environment, the font used is Helvetica 8\,pt, as required by \emph{Journal of the American Chemical Society}.
%The surrounding frame is 9\,cm by 3.5\,cm, which is the maximum permitted for  \emph{Journal of the American Chemical Society} graphical table of content entries. The box will not resize if the content is too big: instead it will overflow the edge of the box.
%This box and the associated title will always be printed on a separate page at the end of the document.
%\end{tocentry}

%\twocolumn[
%\begin{@twocolumnfalse}
\oldmaketitle
\begin{abstract}
With advanced synthetic techniques, a wide variety of well-defined graphene nano\-ribbons (GNRs) can be produced with atomic precision.
Hence, finding the relation between their structures and properties becomes important for the rational design of GNRs.
In this work, we explore the complete chemical space of gulf-edged zigzag graphene nanoribbons (ZGNR-Gs), a subclass of zigzag GNRs in which the zigzag edges miss carbon atoms in a regular sequence.
We demonstrate that the electronic properties of ZGNR-Gs depend on four structural parameters: ribbon width, gulf edge size, unit length, and gulf offset.
Using tight-binding calculations and the Hubbard model, we find that all ZGNR-Gs are semiconductors with varying band gaps; there are no metals in this class of materials.
Notably, when spin polarization is considered, most ZGNR-Gs exhibit antiferromagnetic behavior, with the spin moments and spin-induced band gap opening being stabilized by longer zigzag segments at the edges.
Furthermore, we provide simple empirical rules that describe the $\mathbb{Z}_2$ topological invariant based on the aforementioned structural parameters.
By analyzing the full chemical space of ZGNR-Gs, we offer insights into the design of GNRs with desired electronic, magnetic, and topological properties for nanoelectronic applications.
\end{abstract}
%\end{@twocolumnfalse}
%]


\tocless\section{Introduction}
Graphene nanoribbons (GNRs) are promising materials for nanoelectronics because they are chemically and mechanically stable and have a large variety of electronic properties, such as a wide range of band gaps, spin-polarized edges, or topologically non-trivial electronic states.\cite{tian2023graphene,saraswat2021materials,kumar2023electronic}
The state-of-the-art technique of bottom-up synthesis\cite{10.1039/D0CS01541E,10.1039/C9QM00519F,chen2020graphene} allows atomically precise fabrication of GNRs from programmed precursors, enabling control over their width and edge topology, including regularly shaped terminations (zigzag\cite{ruffieux2016surface} or armchair\cite{liu2020bottom}) or more complex cove\cite{10.1021/jacs.5b03017,10.1021/jacs.1c09000}, gulf\cite{yang2022solution, liu2020synthetic}, and fjord edges\cite{10.1021/jacs.1c01882,10.1021/jacs.0c07657}.

The structure of GNRs determines their electronic properties.\cite{lee2018coves,cao2017topological}
For instance, as predicted from the tight-binding (TB) model, all zigzag GNRs (ZGNR) are metallic, while armchair GNRs (AGNR) are metallic when their width equals $3p + 2$ ($p\in\mathbb{Z}^+$).\cite{son2006energy,10.1103/PhysRevB.54.17954}
Otherwise, AGNRs are semiconducting with a wide variety of band gap sizes.
It is important to note that in the Hubbard model or first-principles calculations, like density functional theory (DFT), which take into account spin polarization, ZGNRs are known to be antiferromagnetic (AFM).\cite{fujita1996peculiar,yazyev2010emergence,son2006half,dutta2008half}
Opposite spins are localized on each side of the zigzag edges, leading to a band gap opening. 

\begin{figure}[ht!]
    \centering
    \includegraphics[width=0.6\textwidth]{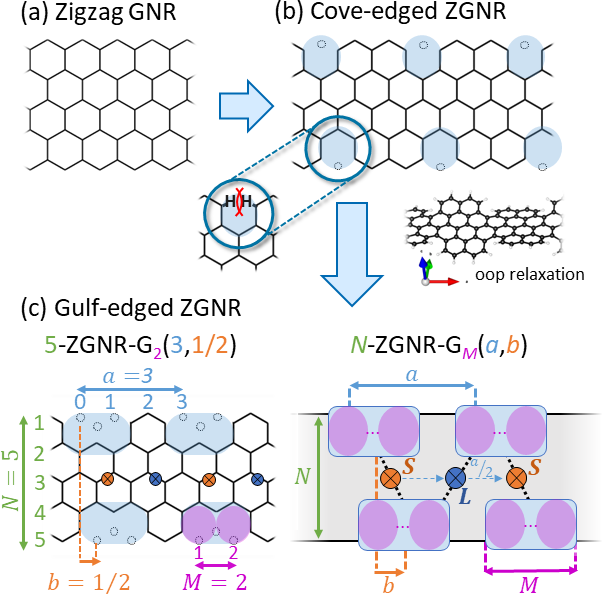}
    \caption{Nomenclature of ZGNR-Gs. The structures of (a) 5-ZGNR, (b) 5-ZGNR-C(3,$\frac{1}{2}$), and (c) 5-ZGNR-G$_2$(3,$\frac{1}{2}$) are given as examples. The parameters $N$, $M$, $a$, and $b$ and the position of reference points \textit{\textbf{S}} and \textit{\textbf{L}} relative to the center of the gulfs are represented schematically. The oop relaxation of the segments in ZGNR-Cs is due to the steric repulsion between the hydrogen atoms inside the cove edges. This effect is absent in ZGNR-Gs. Thus, the ribbons remain planar, similar to ZGNRs.}
    \label{fig_intro_nomenclature}
\end{figure}

In our previous work\cite{arnold2022structure}, we investigated a specific family of GNRs derived from ZGNRs, known as cove-edged ZGNRs (ZGNR-Cs),\cite{lee2018coves,10.1021/jacs.5b03017,10.1021/jacs.1c09000,10.1021/acs.jpcc.0c04152} using the TB model (see Fig.~\ref{fig_intro_nomenclature}(a,b)).
ZGNR-Cs exhibit topologically trivial or non-trivial states, depending on their structural configuration.
In ZGNR-Cs, the hydrogen atoms inside the cove edges introduce a steric hindrance, leading to an out-of-plane (oop) distortion (see Fig.~\ref{fig_intro_nomenclature}(b)).\cite{10.1021/acs.accounts.2c00550,10.1021/jacs.1c09000}
This hampers their on-surface synthesis, a well-suited method for achieving high density and uniformity of various ribbon types.\cite{gu2022nanographenes,saraswat2021materials}
The oop distortion can be avoided by increasing the size of the coves introduced at the edges, resulting in planar, so-called gulf-edged ZGNRs (ZGNR-Gs, in the literature sometimes included into the class of ZGNR-Cs),\cite{liu2020synthetic,yang2022solution,10.1021/acs.accounts.2c00550} shown in Fig.~\ref{fig_intro_nomenclature}(c).
This class of materials has been much less explored than ZGNR-Cs, even though it promises higher compatibility with flat surfaces due to its planar structure and a wider range of tunable properties.

% Figure: visualization differences ZGNR-C vs. ZGNG-G, plus nomenclature of ZGNR-G

In this work, we expand our study from the subset of ZGNR-Cs to ZGNR-Gs as a more general family of edge-patterned ZGNRs.
Structurally unperturbed ZGNRs exhibit AFM ordering, which can be weakened by introducing defects or armchair-like "bites" at the edge.\cite{pizzochero2021edge, kunstmann2011stability, huang2008suppression}.
We explore the magnetic properties of ZGNR-Gs, employing the Hubbard (TB+U) model to examine how the gulf size influences the properties of ZGNR-Gs.
Our findings indicate that significant spin polarization with AFM ordering occurs when the zigzag segments at the edges are sufficiently long, whereas shorter ZGNR-Gs remain diamagnetic. 
Additionally, longer ZGNR-Gs show increased spin polarization energies, localized spin moments, and spin-induced band gap opening. 
The band gap opening leads to the conclusion that no metallic systems exist across the entire ZGNR-G family.
Our TB calculations (corresponding to $U=0$) further show that the $\mathbb{Z}_2$ topological invariant, describing the topological properties of ZGNR-Gs, depends on basic structural parameters and can be described by simple empirical rules.

\tocless\section{Material and methods}

\tocless\subsection{Nomenclature}
To establish a clear scope for our analysis, we only consider ZGNR-Gs with an equal number of \ch{CH} groups removed on both sides of the ribbon.
These symmetric ribbons are more realistic for programmed synthesis as they can be formed from identical building blocks.
As shown in Fig.~\ref{fig_intro_nomenclature}(c), four structural parameters define all ZGNR-G: width $N$, unit length $a$ (in units of hexagonal rings), gulf (cove in the subset of ZGNR-Cs) position offset $b$ on opposite sides of the ribbon, and the gulf size $M$ ($M\in\mathbb{N}, 1\leq M < a$).
The number of connected \ch{C} atoms removed on each side of the ribbon is given by $2M-1$.
$N$-ZGNR-G$_M$($a,b$) then labels all structurally possible regular ZGNR-Gs (see right side of Fig.~\ref{fig_intro_nomenclature}(c); for more details, see the Supplementary Material (SM), section~\ref{sec_SI_nomenclature}).
In this terminology, ZGNR-Cs are a limiting case of ZGNR-Gs with $M = 1$.\cite{arnold2022structure}
However, they are excluded from a detailed discussion, as we focus primarily on ideally flat structures.

ZGNR-Gs contain two distinct inversion centers if $b\neq\frac{a}{2}$.
They are denoted as \textit{\textbf{S}} and \textit{\textbf{L}} in Fig.~\ref{fig_intro_nomenclature}(c) and serve as reference points: \textit{\textbf{S}} is located at the shorter and \textit{\textbf{L}} at the longer offset between adjacent gulf edges.
\textit{\textbf{S}} and \textit{\textbf{L}} are separated by half a lattice vector.
For the exceptional case of $b=\frac{a}{2}$ with Frieze group $p2mg$, all the offsets are equal, resulting in equivalent \textit{\textbf{S}} and \textit{\textbf{L}}.
In our calculations, we choose unit cells with maximum symmetry, so one of the inversion points is at the boundary, and the other is at the center.
This convention is useful for the discussion of topological properties.
As the ribbons are finite in reality, different choices of the unit cells reflect different GNR termination and result in nonequivalent structures, impacting the topological properties as discussed previously\cite{arnold2022structure}.

\tocless\subsection{Computational Details}
The atomistic structure of ZGNR-Gs was created using the ZGNR-Builder tool, available on GitHub.\cite{ZGNRbuilder}

%% Part 1: TB
TB calculations were performed using the PythTB package\cite{PythTB,arnold2022structure} (see SM, section~\ref{sec_SI_TB}, for details).
We used only 1\textsuperscript{st}-neighbor interactions to set up our TB model, and as the systems are conjugated, only one $\pi-$electron per carbon atom was considered. 
Hence, the corresponding TB Hamiltonian $H_{\mathrm{TB}}$ can be written as:
\begin{equation}
H_{\mathrm{TB}}=\sum_{i}{\varepsilon_{i}c^{\dagger}_{i}c_{i}}+\sum_{\left\langle{i,j}\right\rangle}{t_{1}c^{\dagger}_{i}c_{j}}
\end{equation}
where $\varepsilon_{i}$ are the on-site energies and $t_1$ is the hopping parameter of the 1\textsuperscript{st}-neighbor interactions; $c^{\dagger}$ and $c$ are creation and annihilation operators.
Within the TB framework, the $\mathbb{Z}_2$ topological invariant was obtained from the Zak phase\cite{10.1103/PhysRevLett.62.2747}, following
\begin{equation}
    \mathbb{Z}_2 = \left( \frac{\gamma}{\pi} \right) \mathrm{mod}\ 2
\end{equation}
where $\gamma$ is the Zak phase.

%% Part 2: TB+U
To investigate the magnetic properties of ZGNR-Gs, we employed the Hubbard mo\-del,\cite{yazyev2010emergence, ghasemifard2020circumferential} which introduces the repulsive on-site Coulomb interaction as an additional term, $H_{\mathrm{U}}$. 
The resulting Hamiltonian can be written as:
\begin{equation}
H = H_{\mathrm{TB}}+H_{\mathrm{U}} \ \mathrm{with} \ H_{\mathrm{U}}=U\sum_{i}{n_{i\uparrow}n_{i\downarrow}}
\end{equation}
where $U$ is the Hubbard parameter, expressing the magnitude of the Coulomb repulsion, and $n_{i\sigma}=c^{\dagger}_{i\sigma}c_{i\sigma}$ is the spin-resolved electron density at site $i$ for spin channel $\sigma$. 
A mean-field approximation is applied, and the final Hamiltonian can be expressed as:
\begin{equation}
H=H_{\mathrm{TB}}+U\sum_{i}{\left(n_{i\uparrow}\langle n_{i\downarrow}\rangle +\langle n_{i\uparrow}\rangle n_{i\downarrow}-\langle n_{i\uparrow}\rangle \langle n_{i\downarrow}\rangle\right)}.
\end{equation}
Here, $\langle n_{i\sigma}\rangle$ is the expectation value of $n_{i\sigma}$, allowing the transformation to a single-particle picture, where an electron of spin $\sigma$ interacts with the average density of opposite spin at the same site $i$.
The term $-\langle n_{i\uparrow}\rangle \langle n_{i\downarrow}\rangle$ is a correction to avoid double-counting of the interaction energy.
When analyzing ZGNR-Gs, we define them as showing a strong spin polarization once their maximum spin moment reaches at least 50\% of the maximum spin moment in ZGNR systems.

%% Part 3: DFT and GW (very short)
We fitted the TB hopping parameter $t_1$ and Hubbard parameter $U$ to the band gap values obtained from DFT/HSE06\cite{krukau2006influence,heyd2003hybrid} calculations (with tight tier 1 basis\cite{blum2009ab} and a $k$-grid of 18x3x3) as implemented in FHI-aims\cite{blum2009ab, ren2012resolution} using the same geometries as for the TB calculations.
The description of the fitting procedure is given in the SM (see section~\ref{sec_SI_TBU_param}).
The SM also provides more detailed information on the choice of the HSE06 functional based on $GW$ reference calculations (see section~\ref{sec_SI_GW}).
The resulting values are $t_1=3.328$\,eV and $U=1.72t_1=5.723$\,eV (95\% confidence interval: $U=1.65\dots1.79t_1$).
These agree well with values previously reported for conjugated carbon systems\cite{honet2021semi,meena2022ground,schuler2013optimal} and reproduce DFT results well (see SM, section~\ref{sec_SI_TBU_benchmarks}).
%Fig.~\ref{fig_SI_TB+U_benchmark}, \ref{fig_SI_TB+U_benchmark_ZGNR}, and \ref{fig_SI_TB+U_benchmark_bands}).

All calculated data are available as a ZENODO repository.\cite{zenodo}

\tocless\section{Results and discussion}

\tocless\subsection{Electronic structure}
We investigate the electronic structure of ZGNR-Gs and the differences between their diamagnetic and AFM states.
For this, we first study the diamagnetic state using the TB model without incorporating spin polarization (equivalent to $U=0$).
All ZGNR-Gs with even $N$ and $p2mg$ symmetry ($b=\frac{a}{2}$) are metallic, while all the other ZGNR-Gs are semiconducting with a direct band gap at the $\Gamma$ ($\mathrm{X}$) point for odd (even) $a$ (independent of $N$, $M$, and $b$; see Fig.~\ref{fig_SI_bands} for exemplary band structures).
The value of the band gap changes systematically with $b$: for even (odd) $N$, the largest band gaps are observed for $b=0$ ($b=\frac{a}{2}$) and the smallest for $b=\frac{a}{2}$ ($b=0$), resulting in metallic systems for $b=\frac{a}{2}$ if $N$ is even (see Fig.~\ref{fig_SI_table_gap_TB}).
Increasing $N$ and $a$ results in a decreasing the TB band gap as the system converges in the large-scale limit to graphene or ZGNR, respectively, while the band gap increases with $M$, as shown in Fig.~\ref{fig_SI_band_gap_max}.

\begin{figure*}[ht!]
    \centering
    \includegraphics[width=\textwidth]{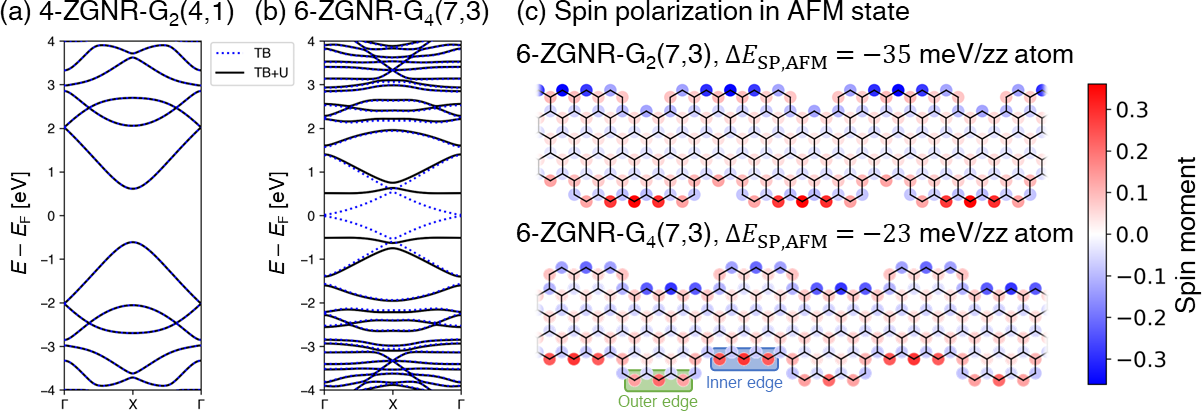}
    \caption{Influence of spin polarization on the electronic structure of exemplary ZGNR-Gs, calculated at TB+U level. (a) In 4-ZGNR-G$_2$(4,1), including spin polarization in the calculation does not influence the band structure. (b) For 6-ZGNR-G$_4$(7,3), the band structure without spin polarization shows a small band gap, while there is a strong band gap opening in the AFM state. Only one spin channel is shown in the AFM state, as both spin channels have identical band structures. Band structures of more systems are given in Fig.~\ref{fig_SI_bands}. (c) Spin polarization plots for the AFM state in 6-ZGNR-G$_2$(7,3) and 6-ZGNR-G$_4$(7,3). The energy difference $\Delta E_\mathrm{SP,AFM}$ per zigzag edge atom between the AFM state and the diamagnetic state is calculated on DFT/HSE06 level. For 6-ZGNR-G$_4$(7,3), the inner and outer zigzag edges are highlighted. Equivalent band structures and spin polarization plots for the AFM and FM state are shown in Fig.~\ref{fig_SI_spin_polarization}.}
    \label{fig_spin_polarization}
\end{figure*}

However, for $U\neq 0$, the band structure near the Fermi level and, thus, the band gap value can change strongly due to spin polarization.
Some ZGNR-Gs are diamagnetic or exhibit only weak magnetic ordering, and their band structures remain unchanged (see Fig.~\ref{fig_spin_polarization}(a)).
In contrast, others host stable AFM states with a band gap opening relative to the diamagnetic state (see Fig.~\ref{fig_spin_polarization}(b)).
In these spin-polarized ZGNR-Gs, sublattices A and B show spin moments up and down, respectively, due to the Ovchinnikov rule\cite{ovchinnikov1978multiplicity}.
Opposite spins are localized on the opposite edges, as shown in Fig.~\ref{fig_spin_polarization}(c), resulting in the overall AFM configuration with the largest spin moments at the zigzag edge atoms.
Throughout the family of ZGNR-Gs, we never observed a ferromagnetic (FM) configuration as the magnetic ground state.
This antisymmetric spin polarization results in a degenerate ground state, which, on the one hand, poses a fundamental issue to DFT.
On the other hand, it makes stabilization of the spin polarization without external symmetry breaking experimentally challenging.\cite{yang2024topological}

% Figure: influence of spin polarization

To gain insight into the AFM states, we define two energy differences: the spin polarization energy of the AFM state $\Delta E_\mathrm{SP,AFM}$ and the energy difference between the different magnetic states $\Delta E_\mathrm{mag}$.
$\Delta E_\mathrm{SP,AFM}$ ($\Delta E_\mathrm{mag}$) is calculated as the energy difference between the AFM state and the diamagnetic (FM) state.
Both are normalized to the number of zigzag (zz) atoms per unit cell $N_\mathrm{zz}$, given by $2(a-1)$.
These quantities, however, cannot be obtained within the TB+U model, so they were thus calculated using DFT, using only a smaller subset of ZGNR-Gs because of the computational cost.
All other properties were calculated using the TB+U model for the whole set of structures.
The value of $\Delta E_\mathrm{SP,AFM}$ is mainly affected by the length of the zigzag segment at the edges.
It decreases with increasing $a$, reaching values down to -47\,meV per zz atom.
For comparison, $\Delta E_\mathrm{SP,AFM}$ for ZGNRs converges to -113\,meV per zz atom.
The absolute energy differences in ZGNR-Gs are smaller than in ZGNRs due to the structural perturbations introduced as gulfs into the zigzag edges.
$\Delta E_\mathrm{mag}$ is either zero (spin-unpolarized systems) or negative, confirming that the AFM state is the magnetic ground state in spin-polarized ZGNR-Gs.
Detailed information on the energetic properties is given in the SM, section~\ref{sec_SI_energies}.

\begin{figure*}[htb!]
    \centering
    \includegraphics[width=\textwidth]{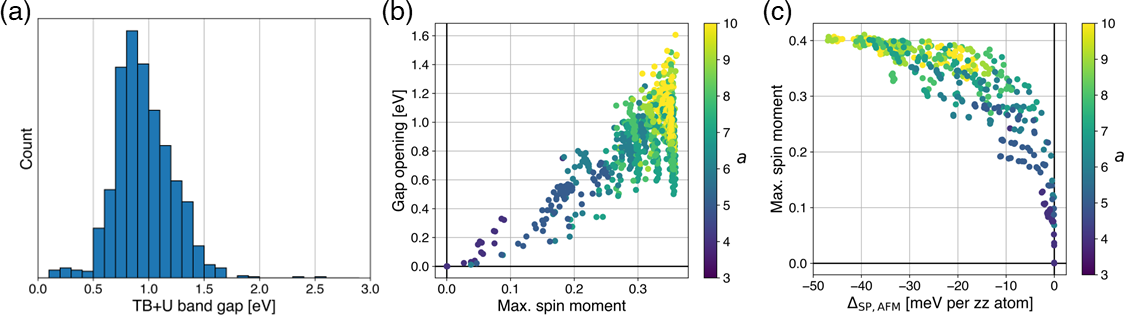}
    \caption{Properties of the AFM state. (a) Distribution of the band gap values, calculated at TB+U level. (b) Band gap opening in the AFM state relative to the diamagnetic state vs. the maximum spin moment, calculated at TB+U level, using the parameter $a$ as colormap. (c) Maximum spin moment vs. $\Delta E_\mathrm{SP,AFM}$, calculated at DFT/HSE06 level, using the parameter $a$ as colormap.}
    \label{fig_properties_AFM}
\end{figure*}

Spin polarization introduces a band gap opening for all systems predicted to be metallic based on the spin-restricted Hamiltonian. 
Therefore, there are no metallic ZGNR-Gs.
Most ZGNR-Gs show band gaps ranging from 0.5 to 1.5\,eV (see Fig.~\ref{fig_properties_AFM}(a). 
Also, within TB+U, the band gaps are always direct.
Due to quantum confinement, the band gap decreases with $N$, the same as the diamagnetic state (see Fig.~\ref{fig_SI_band_gap_vs_a_N}).
As shown in Fig.~\ref{fig_properties_AFM}(b), both band gap opening and the maximum spin moment become more prominent in systems with large $a$, when the zigzag segments at the edges become longer.
However, compared to the diamagnetic state, the strong dependency of the band gap value on $b$ and the alternation of the band gap position with $a$ vanish in the AFM state.
We show detailed information on the influence of structural parameters on band gap size (Fig.~\ref{fig_SI_table_gap_size}), band gap opening (Fig.~\ref{fig_SI_table_gap_opening}), and its position (Fig.~\ref{fig_SI_table_gap_position}) in the SM.

\begin{figure*}[ht!]
    \centering
    \includegraphics[width=0.5\textwidth]{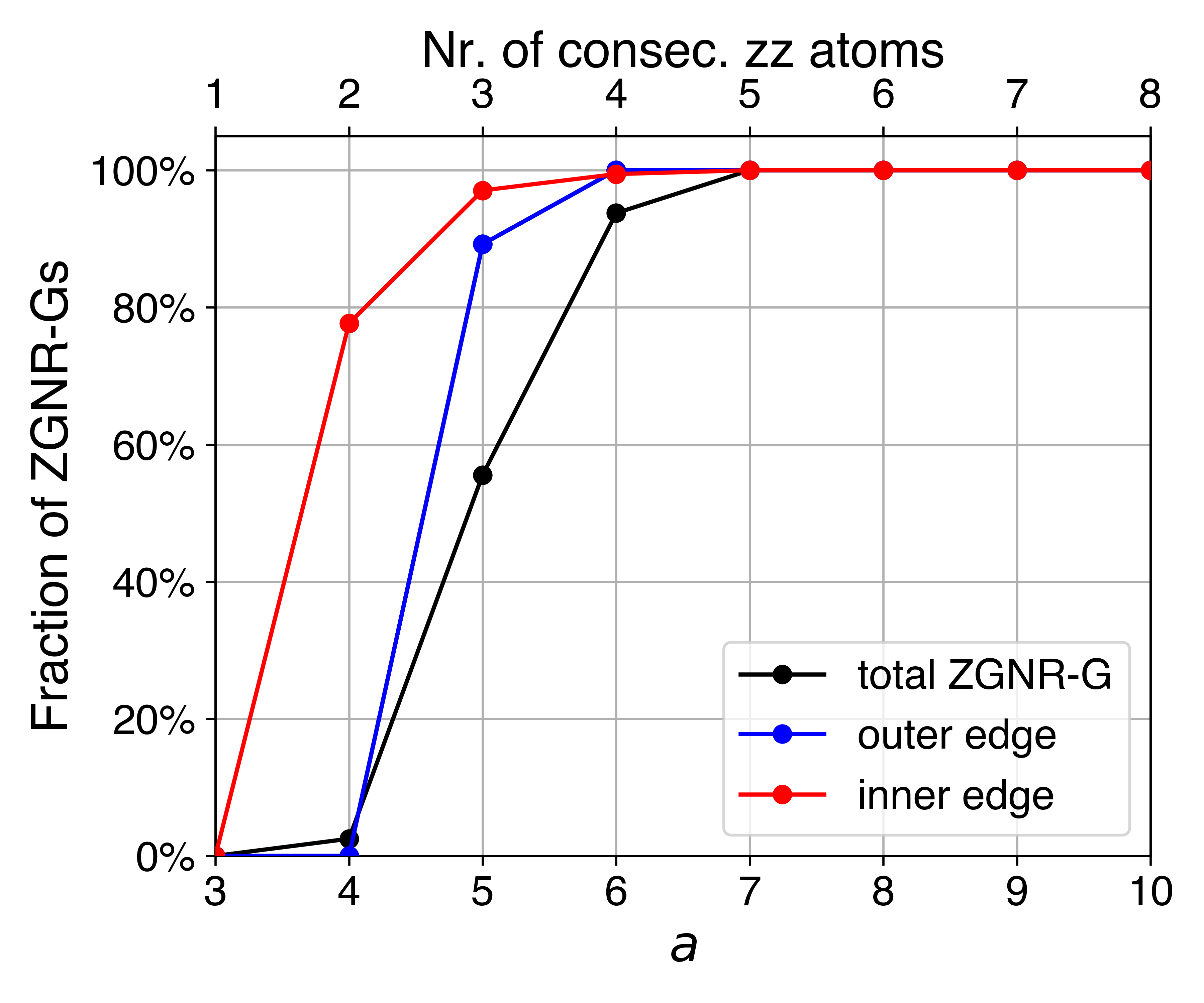}
    \caption{Dependency of the spin polarization on structural parameters. The fraction of the ZGNR-G systems with a strong AFM state at TB+U level is given as function of the parameter $a$ (primary x-axis) and the number of consecutive zigzag edge atoms on the outer and inner edge (secondary x-axis). We define the AFM state as strong once the maximum spin moment reaches at least 50\% of the maximum spin moment in ZGNR-G systems. The equivalent statistics for the calculations of the parametrization set at the DFT/HSE06 level are displayed in Fig.~\ref{fig_SI_AFM_statistics_DFT}, showing good agreement with the TB+U calculations.}
    \label{fig_AFM_statistics_TB+U}
\end{figure*}

As $\Delta E_\mathrm{SP,AFM}$ increases with $a$, the maximum spin moment increases and converges towards the value of the structurally unperturbed zigzag edge in ZGNRs (see Fig.~\ref{fig_properties_AFM}(c); $\approx0.4$ in DFT and $\approx0.35$ in TB+U).
Non-zero spin moments appear for $a\geq 4$, as shown in Fig.~\ref{fig_AFM_statistics_TB+U}.
For $a\geq 6$, ZGNR-Gs typically exhibit a strongly spin-polarized AFM state, corresponding to a continuous zigzag edge segment of 4 atoms or more.
Spin polarization is stronger as $M$ approaches 2 or $a-1$ (see Fig.~\ref{fig_SI_table_spin_max}) when the outer or inner zigzag edge becomes the longest, respectively.
The total spin moment per zigzag edge increases with $a$, giving values up to $\approx0.23$ per zigzag edge atom ($\approx0.33$ per zigzag edge atom in ZGNRs).
We show more details on the spin moments in the SM (see section~\ref{sec_SI_magnetism}).
%(see Fig.~\ref{fig_SI_table_spin_max}, Fig.~\ref{fig_SI_table_spin_average}, and Fig.~\ref{fig_SI_table_spin_per_edge}).
The spin properties only weakly depend on $b$, with the dependency vanishing completely for large $a$. 

Notably, the zigzag segments on the outer and inner parts of the gulf edge behave differently.
As shown in Fig.~\ref{fig_spin_polarization}(c) and supported by Fig.~\ref{fig_SI_outer_vs_inner_edge}, at the same length, the inner zigzag edge shows a larger maximum spin moment than the outer edge.
Furthermore, the outermost zigzag edge atoms on the outer edge show reduced spin moments. 
This difference arises from a partial armchair edge character of the corner atoms compared to a pure zigzag edge character of the central atoms on the outer edge.
Thus, we distinguish inner and outer zigzag edges in the statistics shown in Fig~\ref{fig_AFM_statistics_TB+U}.
To show strong spin polarization, the inner zigzag edge needs at least two consecutive zigzag edge atoms, while the outer edge requires three.
Both edges behave the same for four or more consecutive zigzag edge atoms.

\tocless\subsection{Topological properties}
We now turn to the topological properties of ZGNR-Gs.
In our previous work on ZGNR-Cs, we used the $\mathbb{Z}_2$ topological invariant to classify the topological state.\cite{arnold2022structure}
Including spin polarization in TB+U breaks both the wave function inversion and the time-reversal symmetries, which are necessary for classifying the $\mathbb{Z}_2$ invariant into 0 or 1.
The $\mathbb{Z}$ classification, which relies on chiral symmetry, offers a promising alternative that remains applicable even when time-reversal and spatial symmetries are absent.\cite{jiang2020topology,lopez2021topologically}
However, its applicability to spin-polarized systems has not yet been demonstrated in the literature and is beyond the scope of this work.
Thus, we classify the topological state at the TB level using the $\mathbb{Z}_2$ topological invariant.

\begin{figure*}[ht!] % pagewidth figure
    \centering
    \includegraphics[width=\textwidth]{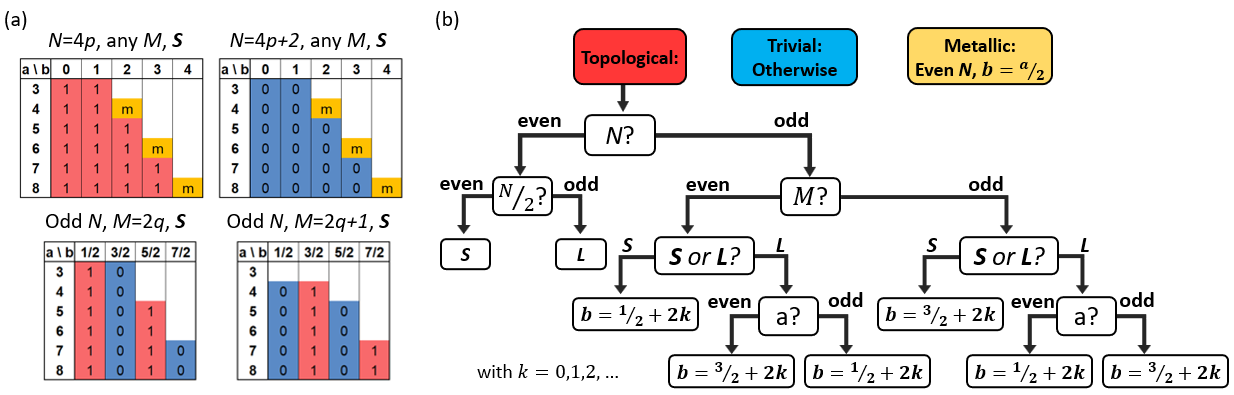}
     \caption{Topological properties of ZGNR-Gs. (a) Table of $\mathbb{Z}_2$ invariant for ZGNR-G, with $\mathbf{S}$ at their boundary, and (b) a scheme for the $\mathbb{Z}_2$ invariant of ZGNR-G depending on the structural parameters width $N$, gulf size $M$, gulf distance $a$, and gulf offset $b$ with integer $p = 1,2,\ldots$ and $q = 1,2,\ldots$.Topological insulators ($\mathbb{Z}_2=1$) are marked in red, trivial semiconductors ($\mathbb{Z}_2=0$) in blue, and metallic ribbons, as obtained in spin-restricted TB calculations, are indicated as "m" (yellow).}
    \label{fig_Z2}
\end{figure*}

%Due to this, we analyzed the chemical space of ZGNR-Gs to better understand their topological properties.
The $\mathbb{Z}_2$ topological invariant in ZGNR-G systems depends not only on the structural parameters but also on the reference points located at the unit cell boundaries.
This is shown in Fig.~\ref{fig_Z2}(a) for unit cells with reference points $\mathbf{S}$ at the boundary (for $\mathbf{L}$ case, see Fig.~\ref{fig_SI_Z2_table_full}).
The general scheme for finding $\mathbb{Z}_2$ of any ZGNR-G, as determined by its four structural parameters $N, M, a,$ and $b$, is shown in Fig.~\ref{fig_Z2}(b).

% Figure: tables of Z2

The $\mathbb{Z}_2$ values of semiconducting ZGNR-Gs without spin polarization depend on $N$:
Independent of $M$, $a$, and $b$, all ZGNR-Gs with even $N$ are topologically nontrivial, if $\frac{N}{2}$ is even (odd) and \textit{\textbf{S}} (\textit{\textbf{L}}) is at the boundary.
For odd $N$ with \textit{\textbf{S}} at the boundary, the $\mathbb{Z}_2$ values switch when $b$ increases incrementally and when $M$ changes between odd and even values. 
With \textit{\textbf{L}} at the boundary, the $\mathbb{Z}_2$ values exhibit a checkerboard pattern, alternating with both $a$ and $b$, as well as between odd and even $M$.
A generalized formula for any ZGNR-Gs (including previously published ZGNR-Cs\cite{arnold2022structure}) can be written to describe all the $\mathbb{Z}^{\boldsymbol{S}}_2$ as the value of $\mathbb{Z}_2$ for reference point \textit{\textbf{S}} as:
\begin{equation}
\begin{split}
    \label{eq_rules_Z2_S}
    \mathbb{Z}^{\boldsymbol{S}}_2= & \left(N+1\right)\mathrm{mod}\ 2 \cdot \left(\frac{N}{2}+1\right)\mathrm{mod}\ 2 \\
    & + \left(N\right)\mathrm{mod}\ 2 \cdot \left(b+\frac{1}{2}+M\right)\mathrm{mod}\ 2
\end{split}
\end{equation}
A function relating $\mathbb{Z}^{\boldsymbol{S}}_2$ to $\mathbb{Z}^{\boldsymbol{L}}_2$ ($\mathbb{Z}_2$ for reference point \textit{\textbf{L}}) is given as:
\begin{equation}
    \label{eq_rules_Z2_L}
    \mathbb{Z}^{\boldsymbol{L}}_2=|\left(Na+1\right)\mathrm{mod}\ 2- \mathbb{Z}^{\boldsymbol{S}}_2|
\end{equation}
We previously rationalized eq.~\ref{eq_rules_Z2_L} using the parity of the involved states.\cite{arnold2022structure}

This demonstrates the versatility of ZGNR-Gs in creating topologically trivial and non-trivial systems.
Since spin polarization can induce a topological phase transition, further investigations, including a topological classification that explicitly considers spin polarization, will be needed to advance our findings further.

\tocless\section{Conclusions}
We provide empirical rules for assessing the electronic properties of the complete chemical space of ZGNR-Gs using the tight-binding method.
These rules are based on the structural parameters -- width ($N$), gulf size ($M$), unit length ($a$), and gulf offset ($b$).
ZGNR-Gs with varying gulf edge sizes behave remarkably similar to the previously reported ZGNR-Cs (with $M=1$), indicating the universality of the governing principles in this class of materials.

ZGNR-Gs exhibit a wide range of band gaps controlled by the choice of structural parameters.
When spin polarization is considered with the Hubbard model, some ZGNR-Gs become antiferromagnetic.
The sublattices show opposite spin moments, particularly prominent at the zigzag edges.
Longer zigzag segments at the edge stabilize the AFM state, increasing spin moment and spin-induced band gap opening.
No metallic systems are present in the whole ZGNR-G family. 
The $\mathbb{Z}_2$ topological invariant, calculated without including spin polarization, changes with $N$, $M$, $a$, and $b$, as well as the choice of the boundary, which can be expressed by simple empirical rules.

These insights allow for the tailored design of ZGNR-Gs with specific electronic and magnetic properties, expanding the potential for creating novel materials with desired functionalities.

\vspace{1cm}
\noindent
\textbf{\Large{Acknowledgement}}\\
The authors thank Deutsche Forschungsgemeinschaft (DFG) for funding within CRC 1415 and SPP 2244 and the European Union’s Horizon 2020 research and innovation program under grant agreement No 956813 for funding this project. D.G. acknowledges the DFG for funding within the Emmy Noether Programme (project number 453275048).
The authors gratefully acknowledge the computing time provided to them on the high-performance computers Barnard and Noctua 2 at the NHR Centers NHR@TUD (ZIH) and NHR@Padeborn (PC\textsuperscript{2}).
This is funded by the Federal Ministry of Education and Research and the state governments participating on the basis of the resolutions of the GWK for the national high-performance computing at universities (www.nhr-verein.de/unsere-partner).
The authors further thank Dr. Thomas Brumme, Beatriz Costa Guedes, and Dr. Hongde Yu for helpful discussions.

%%%%%%%%%%%%%%%%%%%%%%%%%%%%%%%%%%%%%%%%%
%% References moved here before SI starts
%%%%%%%%%%%%%%%%%%%%%%%%%%%%%%%%%%%%%%%%%
\newpage
\renewcommand{\bibsection}{} % no title above bibliography
\noindent
\textbf{\Large{References}}
\providecommand{\latin}[1]{#1}
\makeatletter
\providecommand{\doi}
  {\begingroup\let\do\@makeother\dospecials
  \catcode`\{=1 \catcode`\}=2 \doi@aux}
\providecommand{\doi@aux}[1]{\endgroup\texttt{#1}}
\makeatother
\providecommand*\mcitethebibliography{\thebibliography}
\csname @ifundefined\endcsname{endmcitethebibliography}
  {\let\endmcitethebibliography\endthebibliography}{}

% === Appendices ===
% Specify following sections as appendices (go on then with normal \secion{} commands): \appendix
% Use \appendix* if there is only one appendix
\clearpage
%\onecolumngrid
%\flushleft
\part*{Supplementary Material}

%reset counters in SI
\setcounter{table}{0}
\setcounter{figure}{0}
\setcounter{section}{0}
\setcounter{equation}{0}
%redefining labels to SX
\renewcommand{\thetable}{S\arabic{table}}
\renewcommand{\thefigure}{S\arabic{figure}}
\renewcommand{\thesection}{S\arabic{section}}
\renewcommand{\theequation}{S\arabic{equation}}

\tableofcontents

\newpage
\section{Nomenclature of ZGNR-Gs}
\label{sec_SI_nomenclature}
As shown in Fig.~\ref{fig_intro_nomenclature}(c) of the main text, four structural parameters can unambiguously describe a ZGNR-Gs structure as $N$-ZGNR-G$_M$($a,b$):
\begin{itemize}
    \item $N$: ribbon width, given in terms of carbon zigzag rows ($N\in \mathbb{N}, N\geq 4$),
    \item $M$: size of the gulf edges, given in units of hexagonal rings ($M\in\mathbb{N}, 1\leq M < a$),
    \item $a$: lattice vector where gulf edges reappear on the same side of the ribbon, given in units of hexagonal rings ($a\in\mathbb{N}, a\geq 2$), and
    \item $b$: shortest offset between the centers of adjacent gulf edges on opposite sides of the ribbon ($b\in\left[ 0, \frac{a}{2} \right]$). For even (odd) values of $N$, $b$ is an integer (a half-integer).
\end{itemize}
For any ZGNR-G, the distance between the gulfs $a$ must be larger than the gulf size $M$, requiring $a\geq M+1$.
Meaningful widths are defined for $N\geq4$, as $N=3$ can result in ribbons consisting of units of polycyclic aromatic hydrocarbon, such as naphthalene or biphenyl, connected by \textit{trans}-polyacetylene chains.

\clearpage
\section{Computational details}

\subsection{TB calculations}
\label{sec_SI_TB}
The $\mathbb{Z}_2$ topological invariant is obtained using the Zak phase $\gamma$.
It is calculated from the periodic part of the Bloch function $u_{n\boldsymbol{k}}$ by integrating the Berry connection $i\braket{u_{n\boldsymbol{k}}|\nabla_{\boldsymbol{k}} u_{n\boldsymbol{k}}}$ along an open path $C$ over the Brillouin zone (BZ) and then summing over all occupied states,
\begin{equation}
\gamma=i\sum_{n=1}^{\text{occ.}}\int_C\braket{u_{n\boldsymbol{k}}|\nabla_{\boldsymbol{k}} u_{n\boldsymbol{k}}}\cdot d\boldsymbol{k}.
\end{equation}
The gauge-periodic-boundary condition
\begin{equation}
u_{n\boldsymbol{k}_\mathrm{final}}=e^{-i\boldsymbol{G}\boldsymbol{r}}u_{n\boldsymbol{k}_\mathrm{initial}}    
\end{equation}
is used to close the open path by connecting the cell's last $k$-point with the next cell's first point.\cite{resta2000manifestations}
$\boldsymbol{G}$ is the lattice vector of the reciprocal lattice.
If the system's unit cell has inversion or mirror symmetry, the Zak phase is quantized to zero or $\pi$.
Additionally, one of the inversion centers and the real-space coordinate origin have to coincide to cancel out the intracellular component of the Zak phase.\cite{rhim2017bulk}
Then, the $\mathbb{Z}_2$ topological invariant is calculated from
\begin{equation}
    \mathbb{Z}_2 = \left( \frac{\gamma}{\pi} \right) \mathrm{mod}\ 2.
\end{equation}

Another option to access $\mathbb{Z}_2$ is to use the parity method.\cite{10.1103/PhysRevB.76.045302}
Using the parity $\xi(\psi_n)$ of the $n$ occupied states at the time-reversal invariant momentum (TRIM) points $m$, the topological invariant is obtained from
\begin{equation}
    (-1)^{\mathbb{Z}_2} = \prod_m \prod_n^{\text{occ.}} \xi(\psi_n).
\end{equation}
This method also requires the system's unit cell to be inversion symmetric.
The topological properties obtained from the Zak phase and the parity method agree for all our systems.

\clearpage
\subsection{TB+U parametrization}
\label{sec_SI_TBU_param}
\begin{figure*}[ht!]
    \centering
    \includegraphics[width=0.8\textwidth]{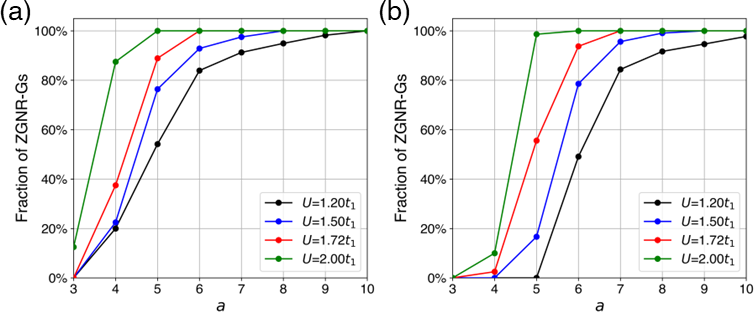}
    \caption{Dependency of the spin polarization on structural parameters. The fraction of the ZGNR systems where the maximum absolute spin moment reaches at least (a) 1\% or (b) 50\% of the maximum spin moment of ZGNR systems is given as function of the parameter $a$ for different values of $U$.}
    \label{fig_SI_AFM_influence_U}
\end{figure*}

Obtaining a suitable Hubbard parameter $U$ is crucial as it is necessary for the agreement of spin moments between DFT and TB+U (see Fig.~\ref{fig_SI_AFM_influence_U}).
Choosing $U$ too large results in a too-strong spin polarization, while too small $U$ causes systems to stay diamagnetic in TB+U despite being AFM in DFT.
Thus, meaningful spin moments can only be obtained from the TB+U calculations after parametrization of $U$.

Due to the high computational cost of DFT, only a small data set with up to 100 atoms was used for calculations at the DFT/HSE06 level.
The large data set of ZGNR-G structures with $N=4\dots11$ and $a=3\dots10$ was used for calculations with the parameterized TB+U model.
The data set used for DFT calculations consists of 465 ZGNR-G systems, while the large data set includes 1248.
All input structures were created with 1.4\,\AA\ \ch{C-C} and 1.1\,\AA\ \ch{C-H} bond lengths without further geometry optimization.
Collinear spin was used in the DFT reference calculations (details given in the method section of the main text) to simulate the AFM state in the spin-polarized calculations.
Initial spin moments were placed by assigning up (down) components of the spin polarization to the top (bottom) outer zigzag edge atoms.
The value of the initial moments was converged to ensure correct magnetic states were obtained.

\begin{figure*}[ht!] % pagewidth figure
    \centering
    \includegraphics[width=0.7\textwidth]{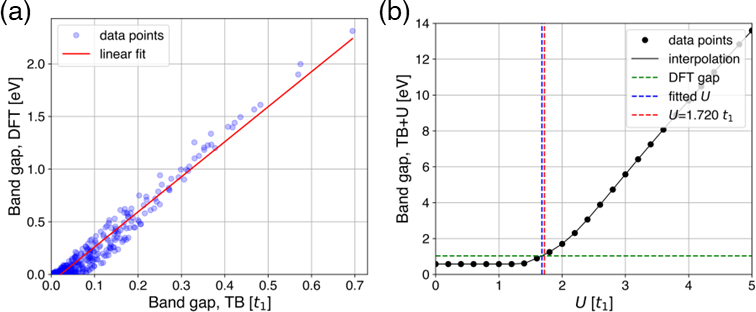}
     \caption{Parametrization of the TB+U model. (a) Results of the $t_1$ parameterization for TB calculations, the linear fit is given by $y=3.328x-0.072$ with $R^2=0.951$. (b) $U$ parameterization for the TB+U calculations demonstrated for $4$-ZGNR-G$_2$(6,0). The value of $U$ is obtained by matching the TB+U band gap to the DFT band gap in the AFM state by linear interpolation between the data points. As a reference, the final $U$ value of 1.720\,eV is shown.}
    \label{fig_SI_parameterization}
\end{figure*}

The hopping parameter for 1\textsuperscript{st}-neighbor interactions $t_1$ was obtained by fitting the band gap obtained in TB calculations against the band gap obtained in DFT calculations without spin polarization.
Systems where the band gap position of DFT was not reproduced in TB were excluded, resulting in 372 usable systems.
We obtained $t_1=3.328$\,eV, as shown in Fig.~\ref{fig_SI_parameterization}(a).

The Hubbard parameter $U$ was obtained from the band gap of the AFM state as calculated on DFT/HSE06 level.
TB+U calculations were performed for each system with the fitted value of $t_1$ obtained above, varying $U$ from $0$ to $5t_1$ with a step size of $0.2t_1$. 
The optimal $U$ value for each structure was then calculated by linear interpolation of the results of the scan of $U$, as shown in Fig.~\ref{fig_SI_parameterization}(b) for an exemplary system.
The final $U$ value was then obtained by averaging over the $U$ values of all structures in the parameterization data set, excluding systems with $U=0$ and those without band gap opening in the TB+U calculation.
The initial spin moments for the TB+U calculations were determined by assigning the atoms in sublattice A (B) an up (down) initial spin moment of 0.2.

\clearpage
\subsection{GW calculations}
\label{sec_SI_GW}
We performed  single-shot $GW$ calculations with the FHI-aims program package~\cite{blum2009ab_SI,Ren2021} using the Perdew-Burke-Ernzerhof (PBE) functional~\cite{Perdew1996}.
The procedure is denoted as $G_0W_0@$PBE.
FHI-aims expands the Kohn-Sham DFT orbitals in numeric atom-centered orbitals (NAOs).
If not otherwise stated, we used NAOs of \textit{tier1} quality.
The periodic $G_0W_0$ implementation in FHI-aims utilizes a local resolution-of-the-identity (RI) approach known as RI-LVL\cite{Ihrig2015,Ren2021} to reformulate the four-center two-electron Coulomb repulsion integrals in a product of two- and three-center Coulomb integrals.
The RI-LVL approach requires auxiliary basis functions (ABFs), which are auto-generated in FHI-aims.
We added hydrogen-like $4f$ and $5g$ ABFs to the auto-generated set to minimize the RI-LVL error.
We analytically continued the $G_0W_0$ self-energy from the imaginary to the real frequency axis using a Pad\'{e} model~\cite{Vidberg1977} with 16 parameters.
We used a modified Gauss-Legendre grid\cite{ren2012resolution_SI} with 60 frequency points to perform the frequency integration of the self-energy.
The quasiparticle equation was solved iteratively.
If not otherwise indicated, we employed an $18\times 3\times 3$ $k$-mesh.
Using more than one $k$-point in the non-periodic directions was necessary due to limitations in the current approach\cite{Ren2021} implemented in FHI-aims to treat the divergence of $GW$ at the $\Gamma$ point.

\begin{figure*}[ht!] % pagewidth figure
    \centering
    \includegraphics[width=\textwidth]{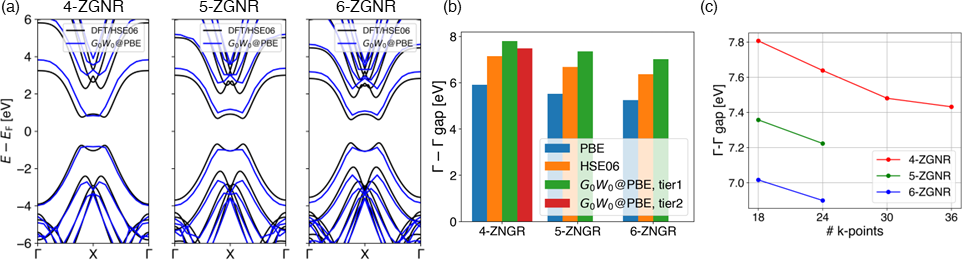}
     \caption{Results of the GW calculations for ZGNRs. (a) Comparison of the band structure of (left to right) 4-ZGNR, 5-ZGNR, and 6-ZGNR, calculated at DFT/HSE06 level and at $G_0W_0$@PBE level with a tight tier 1 basis and a $k$-grid of 18x3x3. (b) Comparison of the $\Gamma-\Gamma$ gaps at the DFT/PBE, DFT/HSE06, and $G_0W_0$@PBE level of theory. (c) Dependency of the $\Gamma-\Gamma$ gap on the number of $k$-points along the GNRs extended length for $G_0W_0$@PBE calculations with a tight tier 1 basis set.}
    \label{fig_SI_GW}
\end{figure*}

$GW$ calculations\cite{Hedin1965,Golze2019} are computationally orders of magnitude more expensive than DFT calculations.
Thus, we restricted our investigation to 4-ZNGR, 5-ZNGR, and 6-ZNGR.
Fig.~\ref{fig_SI_GW}(a) compares the $GW$ to the DFT/HSE06 band structures.
We find that DFT/HSE06 reproduces the $GW$ results generally well.
Fig.~\ref{fig_SI_GW}(b) displays the quantitative comparison of the $\Gamma-\Gamma$ gaps at the DFT/PBE, DFT/HSE06, and $GW$@PBE levels of theory.
DFT/HSE06 opens the gap by 1.1 to 1.2~eV with respect to DFT/PBE.
The $GW$ gaps are still $\approx 0.7$~eV larger than the HSE06 ones.
However, the $GW$ calculations are not fully converged: The convergence of $GW$ is known to be slow with respect to basis set size\cite{Golze2019} and, in the case of 1D or 2D systems, with respect to the $k$-points.\cite{Qiu2016,Golze2019}
We computed the $GW$ gaps for 4-ZNGR also with a larger \textit{tier 2} basis set, which reduces the $GW$ gap by 0.3~eV compared to the results obtained with the \textit{tier 1} basis set (see Fig.~\ref{fig_SI_GW}(b)).
The dependence on the $k$-grids is shown in Fig.~\ref{fig_SI_GW}(c), indicating that 36 $k$-points are required in the periodic direction to converge the calculations, reducing the gap by 0.4~eV compared to calculations with 18 $k$-points.
Considering both convergence trends, the fully converged $\Gamma-\Gamma$ gaps from $GW$ are expected to be $\approx 0.7$~eV smaller, which matches the DFT/HSE06 predictions.

\clearpage
\subsection{TB+U benchmarks}
\label{sec_SI_TBU_benchmarks}
\begin{figure*}[ht!] % pagewidth figure
    \centering
    \includegraphics[width=\textwidth]{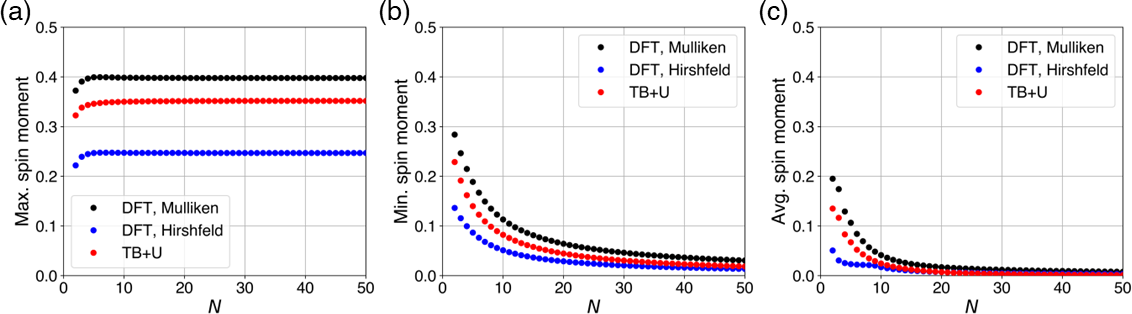}
     \caption{Dependency of the spin polarization in ZGNRs on their width $N$. (a) Maximum, (b) minimum, and (c) average absolute spin moment of the carbon atoms of ZGNR with increasing width $N$, obtained from Mulliken projection and Hirshfeld analysis based on DFT and TB+U calculations.}
    \label{fig_SI_Mulliken_vs_Hirshfeld}
\end{figure*}

The spin moments in the AFM state at the DFT/HSE06 level were obtained using Mulliken projection\cite{mulliken1955electronic}, agreeing well with the values obtained from TB+U, as shown in Fig.~\ref{fig_SI_Mulliken_vs_Hirshfeld} for ZGNRs of different widths $N$.
The maximum value at the zigzag edge for wide ribbons converges to 0.398 in Mulliken analysis using DFT and 0.352 using TB+U.
Using Hirshfeld analysis\cite{hirshfeld1977bonded} instead of the Mulliken analysis does not change the results qualitatively but only gives smaller spin polarization values with a maximum of 0.247 for large $N$.

\begin{figure*}[ht!] % pagewidth figure
    \centering
    \includegraphics[width=\textwidth]{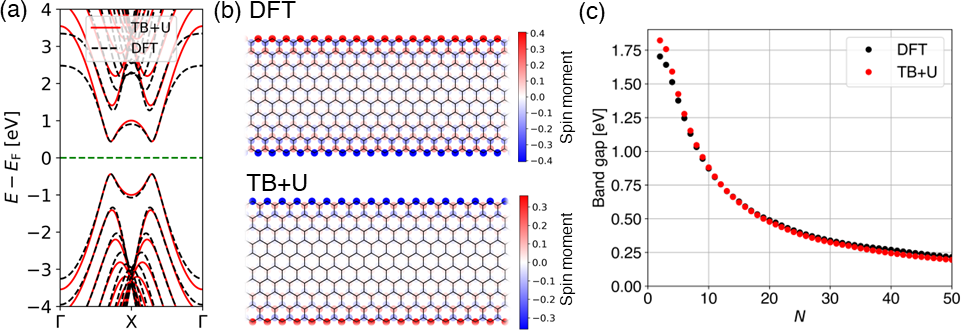}
    \caption{Differences between results obtained from DFT/HSE06 and TB+U for ZGNRs: (a) Overlay of band structures for the AFM state in 10-ZGNR. (b) Comparison of the spin polarization in 10-ZGNR. (c) Band gap in ZGNR as function of their width $N$.}
    \label{fig_SI_TB+U_benchmark_ZGNR}
\end{figure*}

The band structures of ZGNR systems agree well between TB+U and DFT, as shown in Fig.~\ref{fig_SI_TB+U_benchmark_ZGNR}(a).
As visible in Fig.~\ref{fig_SI_Mulliken_vs_Hirshfeld}(b), the decay of the minimum spin polarization in TB+U is faster than in DFT.
This is also visible in Fig~\ref{fig_SI_TB+U_benchmark_ZGNR}(b): the center of the ZGNR system has a non-zero spin polarization in DFT, while it decays much faster with increasing $N$ in TB+U.
However, the band gap of the AFM state shows excellent agreement between TB+U and DFT calculations (see Fig.~\ref{fig_SI_TB+U_benchmark_ZGNR}(c)).

\begin{figure*}[ht!] % pagewidth figure
    \centering
    \includegraphics[width=\textwidth]{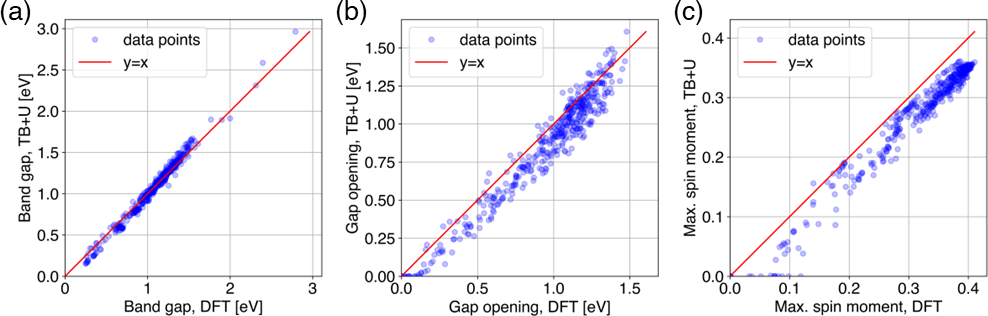}
    \caption{Differences between results obtained from DFT/HSE06 and TB+U for ZGNR-Gs: (a) Band gap of the AFM state, (b) band gap opening relative to the diamagnetic case, and (c) maximum spin moment. In all plots, the red line indicates "y=x" -- the line of perfect agreement between DFT and TB+U calculations.}
    \label{fig_SI_TB+U_benchmark}
\end{figure*}

In ZGNR-Gs, the band gaps of the AFM state $\Delta_\mathrm{TB+U}$ calculated with our parameterized TB+U model agree well with the values $\Delta_\mathrm{DFT}$ obtained from DFT/HSE06 calculations (see Fig.~\ref{fig_SI_TB+U_benchmark}(a)), giving
\begin{equation}
    \Delta_\mathrm{TB+U}=1.1\cdot\Delta_\mathrm{DFT}-0.1, \ R^2=0.986.
\end{equation}
As the main difference, we observe that systems that only show a small band gap opening below 150\,meV, or a weak spin polarization below 20\% of the maximum value in DFT exhibit no AFM state in TB+U but remain diamagnetic (see Fig.~\ref{fig_SI_TB+U_benchmark}(b,c)).

\begin{figure*}[ht!] % pagewidth figure
    \centering
    \includegraphics[width=\textwidth]{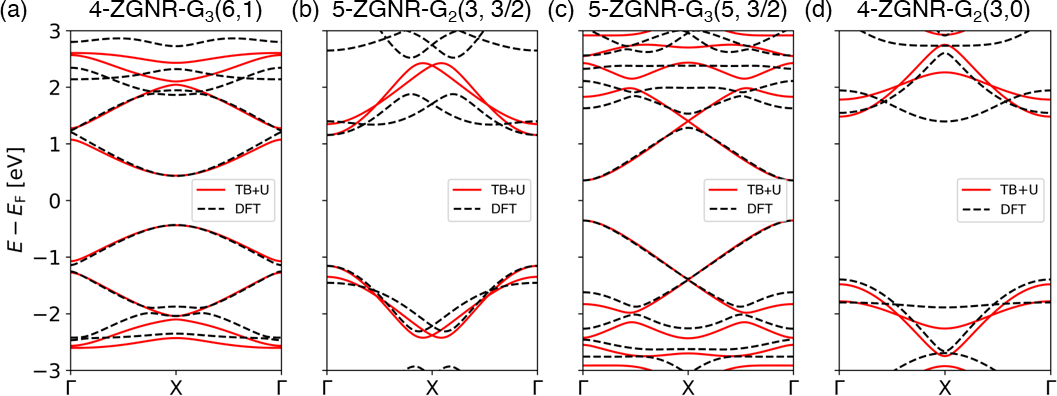}
    \caption{
    Overlay of band structures calculated at the TB+U and DFT/HSE06 level (AFM state) for different exemplary systems.}
    \label{fig_SI_TB+U_benchmark_bands}
\end{figure*}

The band structure of ZGNR-Gs close to the Fermi level typically agrees well between TB+U and DFT (exemplary system shown in Fig~\ref{fig_SI_TB+U_benchmark_bands}(a)).
Bands further away from the Fermi level agree less, as the conjugated $\pi$ system less strongly dominates these bands.

\newpage
However, we observe the following discrepancies for some of the systems:
\begin{itemize}
    \item Strongly varying dispersion in the valence band (VB) or conduction band (CB) (see Fig~\ref{fig_SI_TB+U_benchmark_bands}(b)).
    \item Band crossings between VB/VB-1 or CB/CB+1 which are not present in both TB+U and DFT/HSE06 (see Fig~\ref{fig_SI_TB+U_benchmark_bands}(c)).
    \item Systems with 4-ZGNR-G$_M$($M+1$,0) show an opposite dispersion for the VB and CB between TB+U and DFT. This potentially originates from issues with reproducing an in-plane C-C hybridization in TB+U. This issue is only observed for $N=4$, where the part of GNR inside the gulf edge has a width of $N=2$ and thus locally represents a polyacene structure.
\end{itemize}
The last issue described causes an interesting outlier in the data set: in 4-ZGNR-G$_2$(3,0), the TB+U band gap is direct, while the DFT band gap is indirect.
The difference between the direct and indirect DFT band gap is 150\,meV -- this is the only system where we observe such a strong difference.
Understanding this further is beyond the scope of this work.

\clearpage
\section{Results}

\subsection{Band structure of ZGNR-Gs}
\begin{figure*}[htp!] % pagewidth figure
    \centering
    \includegraphics[width=0.75\textwidth]{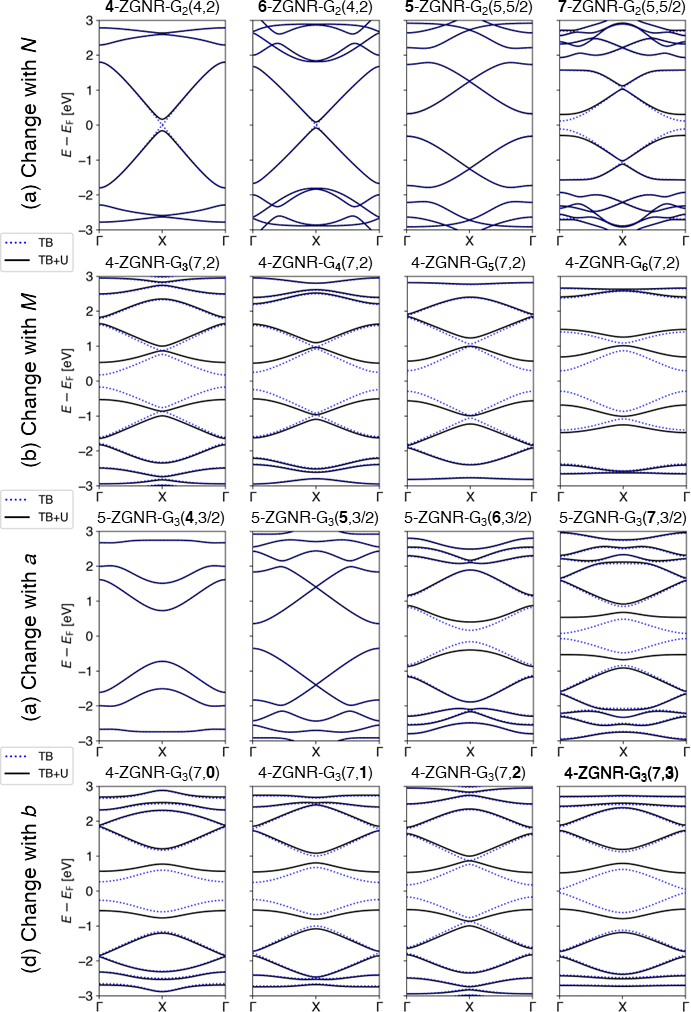}
    \caption{
    Band structures of exemplary ZGNR-G systems, calculated at the TB and TB+U levels of theory, with varying structural parameters: (a) $N$, (b) $M$, (c) $a$, and (d) $b$. Only one spin channel is shown in the AFM state, as both spin channels have identical band structures.}
    \label{fig_SI_bands}
\end{figure*}

\clearpage
\subsection{Electronic properties of the diamagnetic state}
\begin{sidewaysfigure}[ht!]
    \centering
    \includegraphics[width=\textwidth]{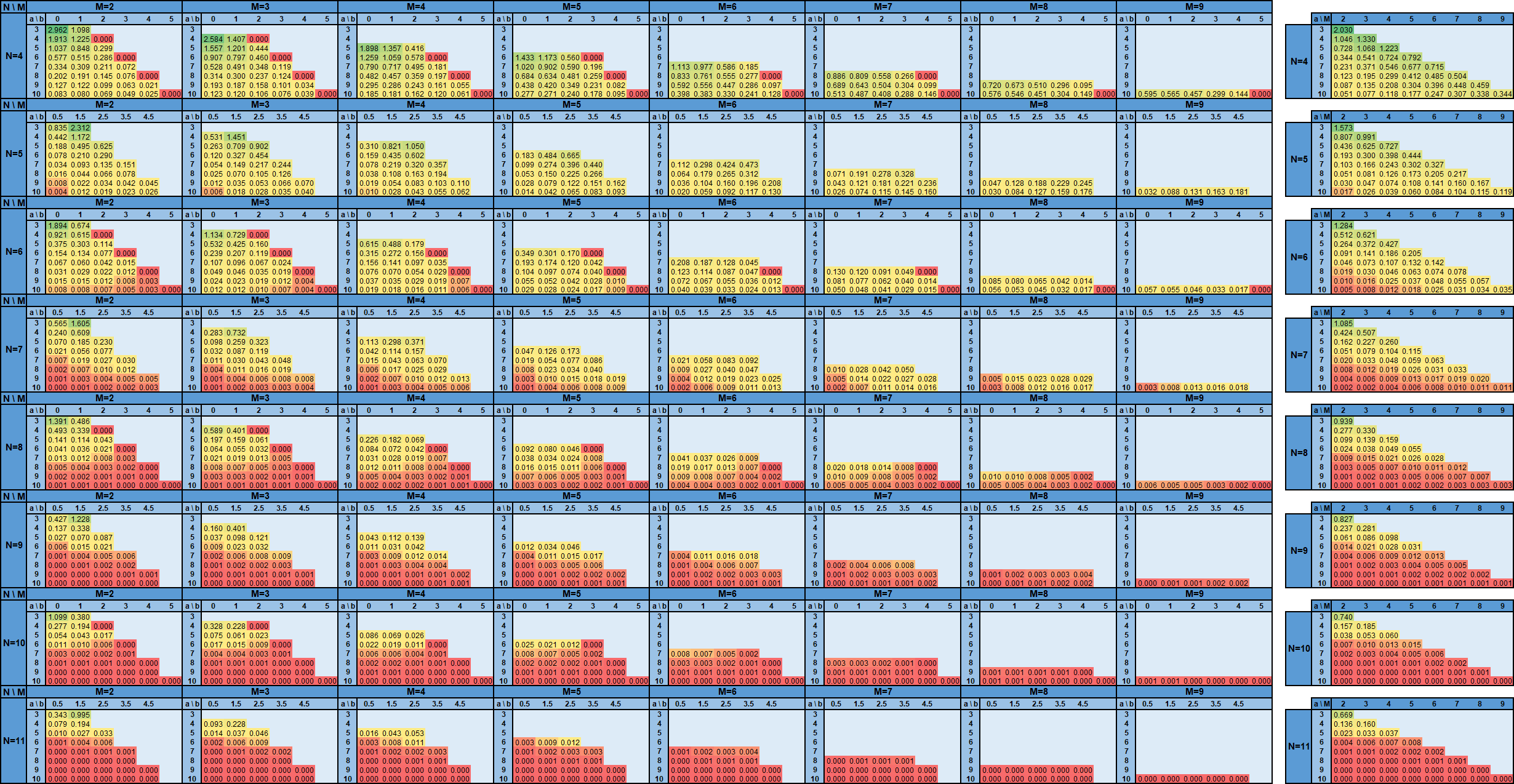}
    \caption{
    Dependency of the band gap (in units of eV) in the diamagnetic state on the structural parameters in ZGNR-Gs, obtained from TB calculations.}
    \label{fig_SI_table_gap_TB}
\end{sidewaysfigure}

The dependency of the band gap in the diamagnetic state, as obtained in spin-restricted TB calculations, on the structural parameters $N$, $M$, $a$, and $b$ is given in Fig.~\ref{fig_SI_table_gap_TB}.
For all $M$, the dependence of the band gap $\Delta_\mathrm{g}$ on $b$ can be estimated as
\begin{equation}
    \Delta_\mathrm{g}= \Delta_\mathrm{max} \cdot \cos \left[\left(\frac{b}{a}+\frac{N\mathrm{mod\ 2}}{2}\right)\pi\right],
    \label{eq_band_gap_gulfs}
\end{equation}
where $\Delta_\mathrm{max}$ is the maximum band gap within a ZGNR-G series with the same $M$, $N$, and $a$.
The band gap dependence on $b$ -- expressed by the cosine term in eq.~\ref{eq_band_gap_gulfs} -- results in metallic systems only for even $N$ with $b=\frac{a}{2}$.
In all other cases, ZGNR-Gs are semiconducting.
With this, ZGNR-Gs behave identically to ZGNR-Cs, where the same dependency of the band gap on $b$ was observed.\cite{arnold2022structure_SI}

\begin{figure*}[ht!] % pagewidth figure
    \centering
    \includegraphics[width=\textwidth]{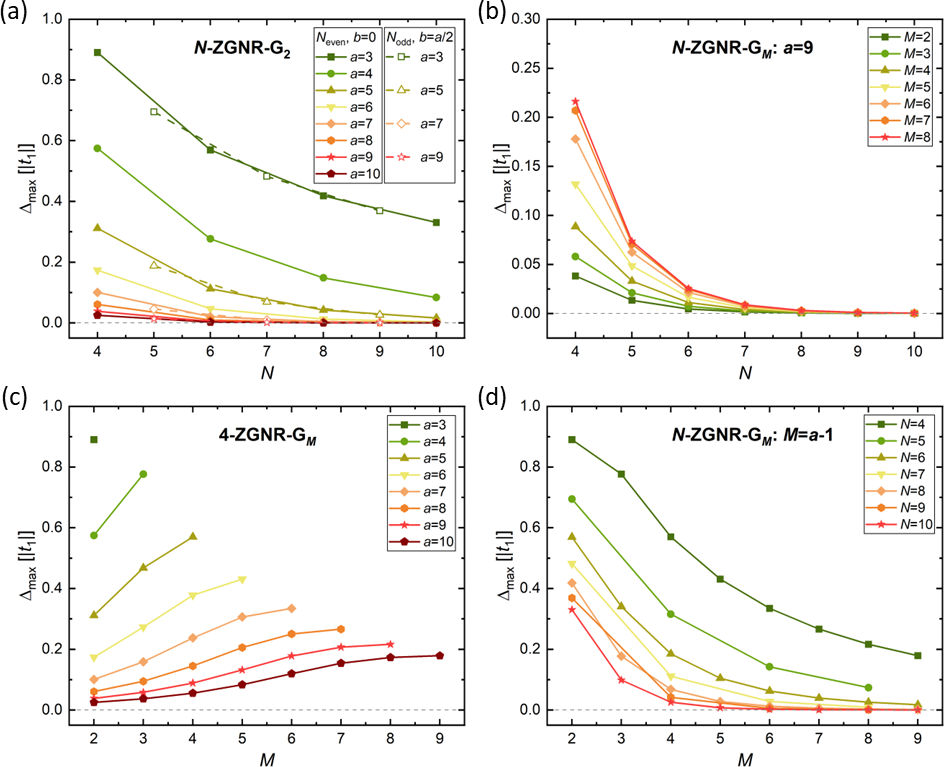}
    \caption{
    Maximum band gaps $\Delta_\mathrm{max}$, calculated at the TB level. Values are given in units of the 1\textsuperscript{st}-neighbor interaction parameter $|t_1|$ as function of (a, b) $N$ and (c, d) $M$, varying (a, c) $a$, (b) $M$, and (d) $N$ between differently colored lines.}
    \label{fig_SI_band_gap_max}
\end{figure*}

The dependence of $\Delta_\mathrm{max}$ on the structural parameters is shown in Fig.~\ref{fig_SI_band_gap_max}.
$\Delta_\mathrm{max}$ decreases with increasing $N$ and $a$ due to quantum confinement, as presented in Fig.~\ref{fig_SI_band_gap_max}(a) with fixed $M = 2$ and Fig.~\ref{fig_SI_band_gap_max}(b) with fixed $a = 9$.
Increasing the gulf size for a given $N$ and $a$ increases $\Delta_\mathrm{max}$. 
We demonstrate this in Fig.~\ref{fig_SI_band_gap_max}(c) for the exemplary case of $N = 4$.
For the largest gulf size, $M=a-1$, ZGNR-Gs can be seen as a pristine ZGNR of width $N-2$ with isolated carbon hexagon rings appearing periodically on both edges.
In this case, $\Delta_\mathrm{max}$ decreases as $M$ increases with fixed $N$ (see Fig.~\ref{fig_SI_band_gap_max}(d)).
This behavior was already described before\cite{cassiano2020smooth}.
%An outlier of this trend is $M=1$, showing a gap much smaller than $M=2$.
%The bands near the Fermi level can change strongly with $M$, but there is no change to the position of the band gap.

% Figure: maximum band gap in ZGNR-G

\clearpage
\subsection{Comparison of AFM and FM states}
\begin{figure*}[ht!]
    \centering
    \includegraphics[width=\textwidth]{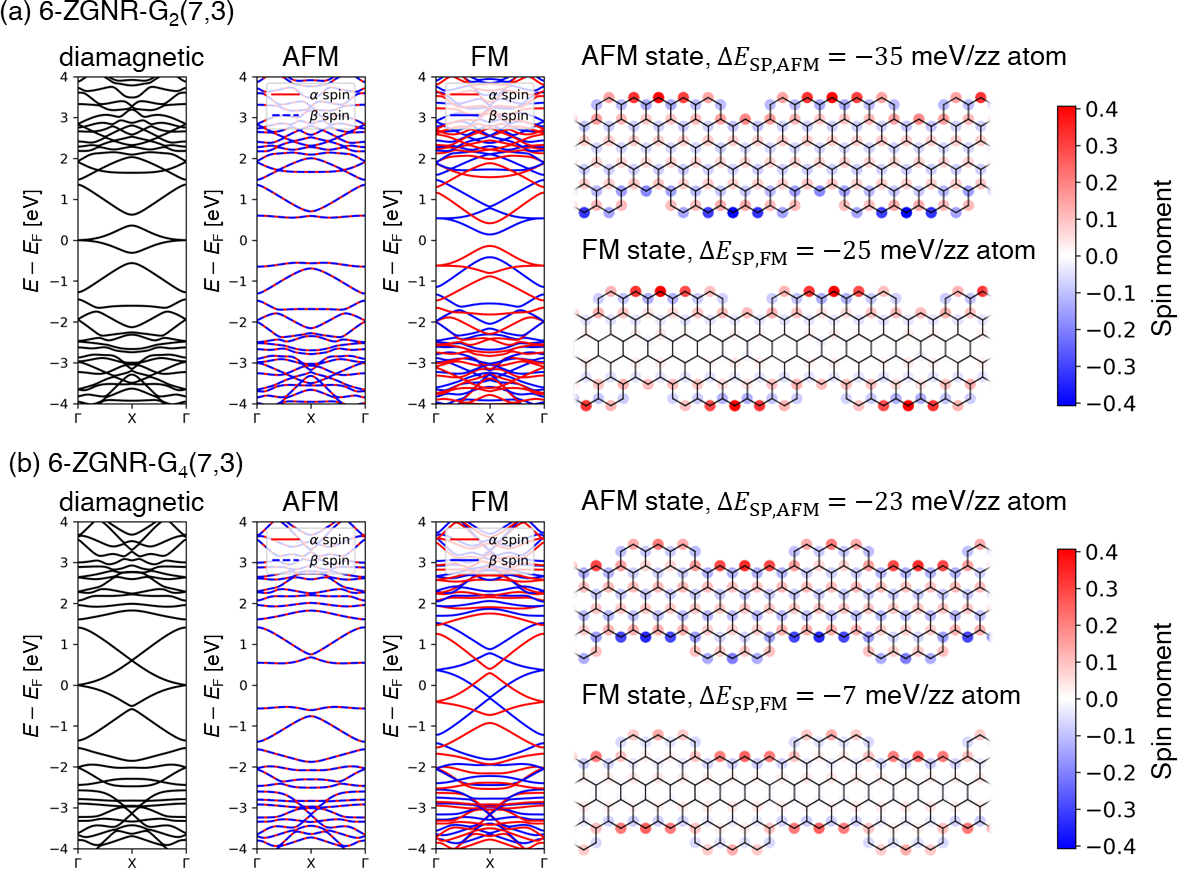}
    \caption{Influence of spin polarization on the band structure of exemplary ZGNR-Gs. Results are calculated at the DFT/HSE06 level for (a) 4-ZGNR-G$_2$(4,1) and (b) 4-ZGNR-G$_4$(7,3). The band structure is shown (left to right) for the diamagnetic, AFM, and FM states. The spin polarization plots are shown for the AFM and FM state. The energy difference $\Delta E_\mathrm{SP,AFM}$ ($\Delta E_\mathrm{SP,FM}$) per zigzag edge atom between the AFM (FM) state and the diamagnetic state is calculated at the DFT/HSE06 level.}
    \label{fig_SI_spin_polarization}
\end{figure*}

\clearpage
\subsection{Energetic properties}
\label{sec_SI_energies}
\begin{figure*}[ht!]
    \centering
    \includegraphics[width=0.5\textwidth]{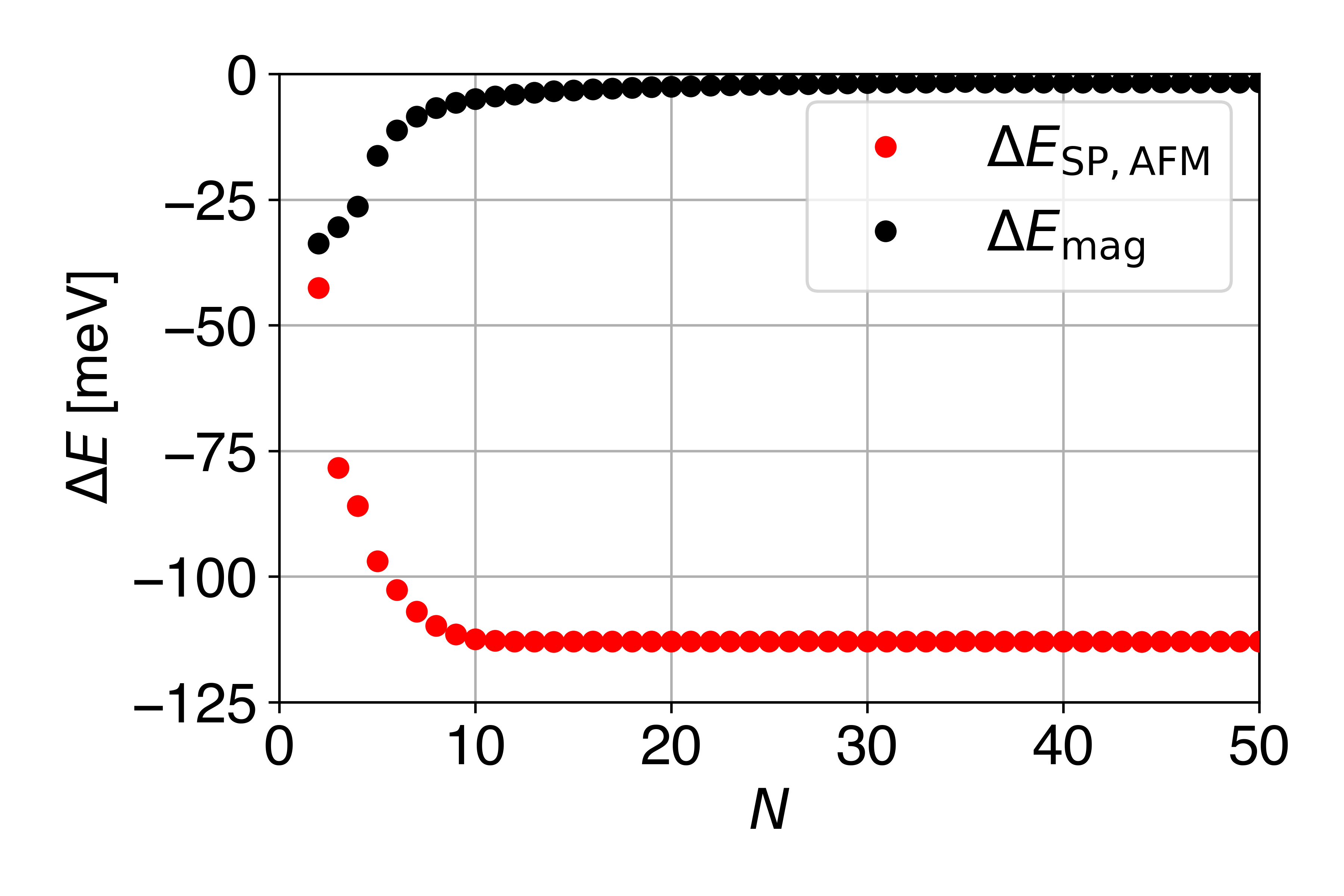}
    \caption{
    Spin polarization energy $\Delta E_\mathrm{SP,AFM}$ and energy difference between the magnetic states $\Delta E_\mathrm{mag}$ for ZGNR systems of different widths $N$, calculated using DFT.}
    \label{fig_SI_energy_ZGNR}
\end{figure*}

\begin{sidewaysfigure}[ht!]
    \centering
    \includegraphics[width=\textwidth]{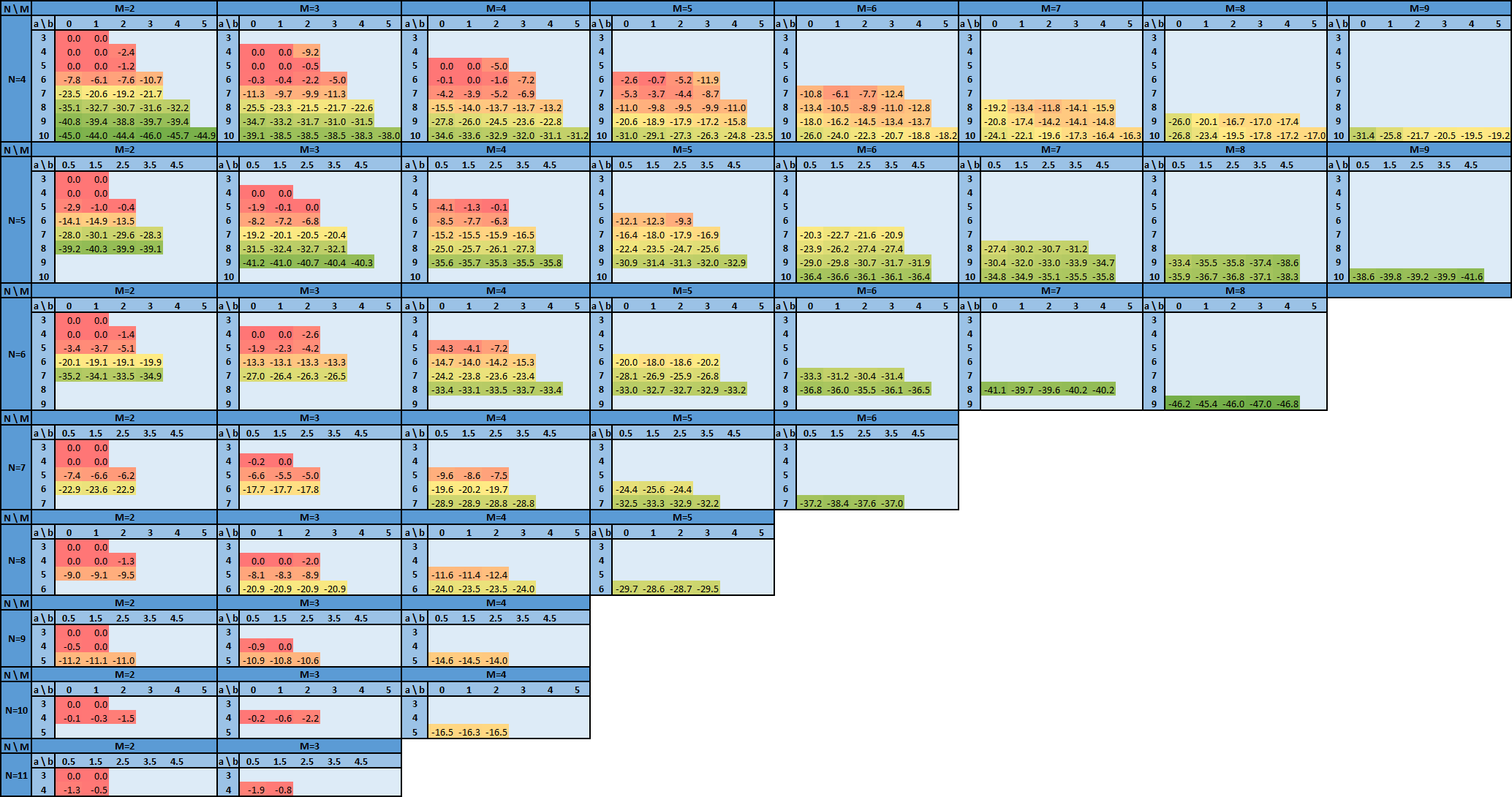}
    \caption{
    Spin polarization energy $\Delta E_\mathrm{SP, AFM}$ of ZGNR-Gs, calculated using DFT.}
    \label{fig_SI_table_energy_AFM_vs_dia}
\end{sidewaysfigure}

\begin{sidewaysfigure}[ht!]
    \centering
    \includegraphics[width=\textwidth]{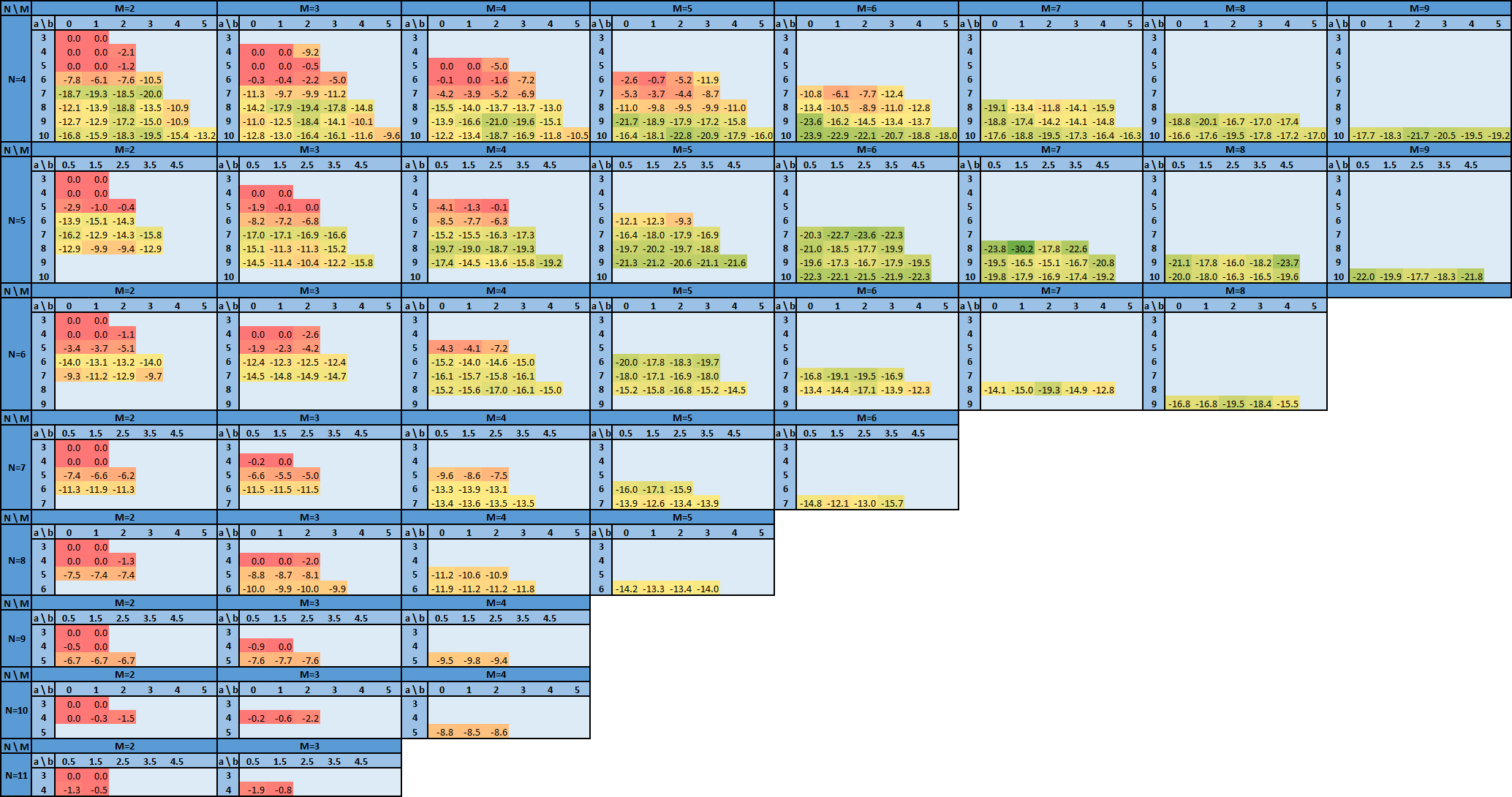}
    \caption{
    Energy difference between different magnetic states $\Delta E_\mathrm{mag}$ of ZGNR-Gs, calculated using DFT.}
    \label{fig_SI_table_energy_AFM_vs_FM}
\end{sidewaysfigure}

The spin polarization energy $\Delta E_\mathrm{SP,AFM}$ is calculated from the energy of the AFM state $E_\mathrm{AFM}$ and the diamagnetic state $E_\mathrm{dia}$ following
\begin{equation}
    \Delta E_\mathrm{SP,AFM}=\frac{E_\mathrm{AFM}-E_\mathrm{dia}}{N_\mathrm{zz}}
\end{equation}
while $\Delta E_\mathrm{mag}$ is calculated between the AFM state and the FM state
\begin{equation}
    \Delta E_\mathrm{mag}=\frac{E_\mathrm{AFM}-E_\mathrm{FM}}{N_\mathrm{zz}}.
\end{equation}
Both quantities are normalized by the number of zigzag edge atoms $N_\mathrm{zz}=2\cdot(a-1)$ to compare different ZGNR-Gs and with ZGNRs as reference systems.
The values for ZGNR systems of different width $N$ are given in Fig.~\ref{fig_SI_energy_ZGNR}.
$\Delta E_\mathrm{mag}$ is converging to $\approx-2$\,meV per zz atom, showing a decreasing difference between the AFM and FM states for large $N$ due to the large distance between the zigzag edges.
The values for ZGNR-Gs are given in Fig.~\ref{fig_SI_table_energy_AFM_vs_dia} and Fig.~\ref{fig_SI_table_energy_AFM_vs_FM}.
All values of $\Delta E_\mathrm{SP,AFM}$ are non-zero for $a\geq7$, showing that all of these systems exhibit an AFM magnetic ground state.

\clearpage
\subsection{Electronic properties of spin-polarized ZGNR-Gs}
\begin{figure*}[ht!]
    \centering
    \includegraphics[width=0.5\textwidth]{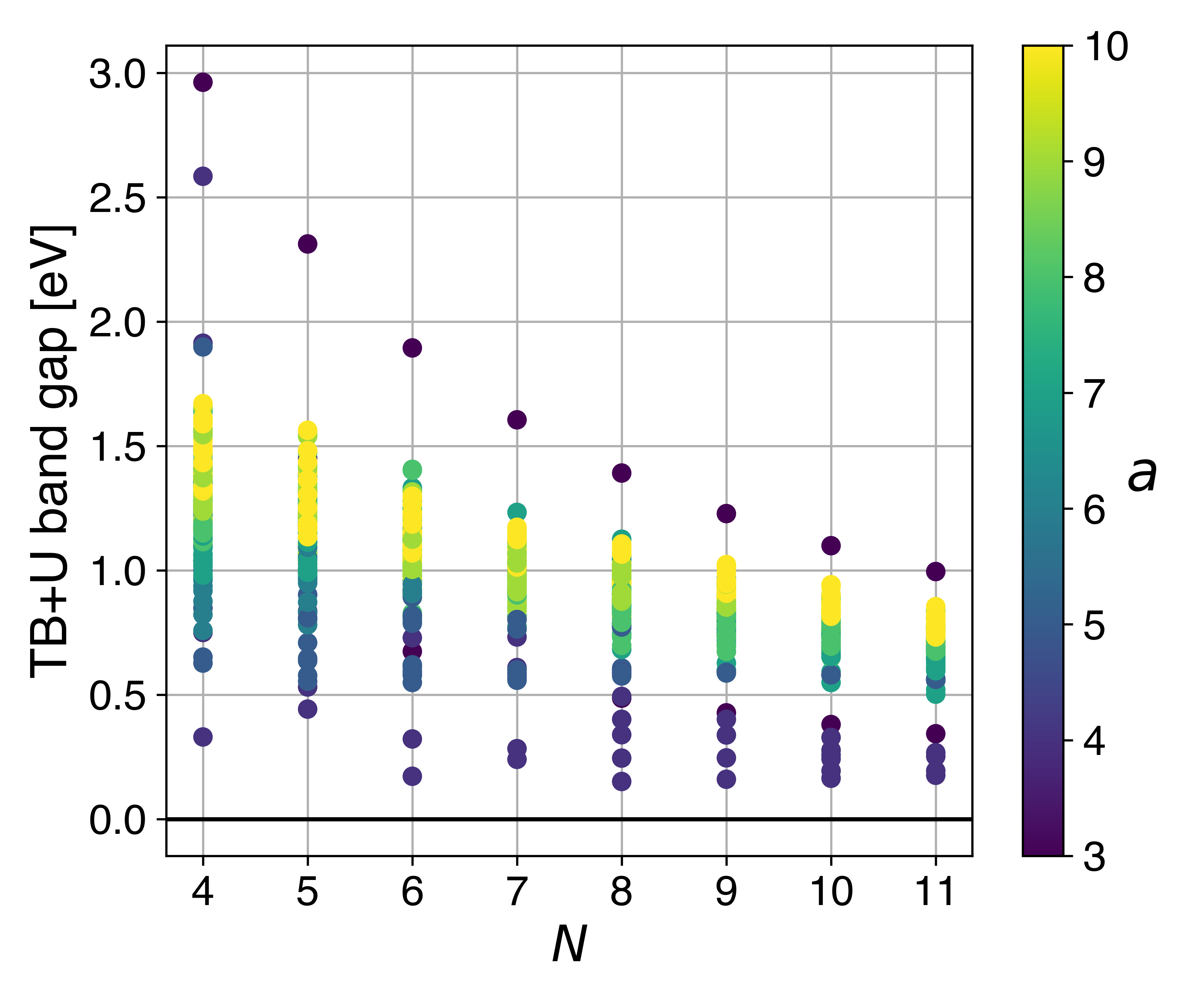}
    \caption{
    Band gap of spin-polarized ZGNR-Gs, calculated on TB+U level, as function of the ZGNR-Gs width $N$, using $a$ as colormap.}
    \label{fig_SI_band_gap_vs_a_N}
\end{figure*}

\begin{sidewaysfigure}[ht!]
    \centering
    \includegraphics[width=\textwidth]{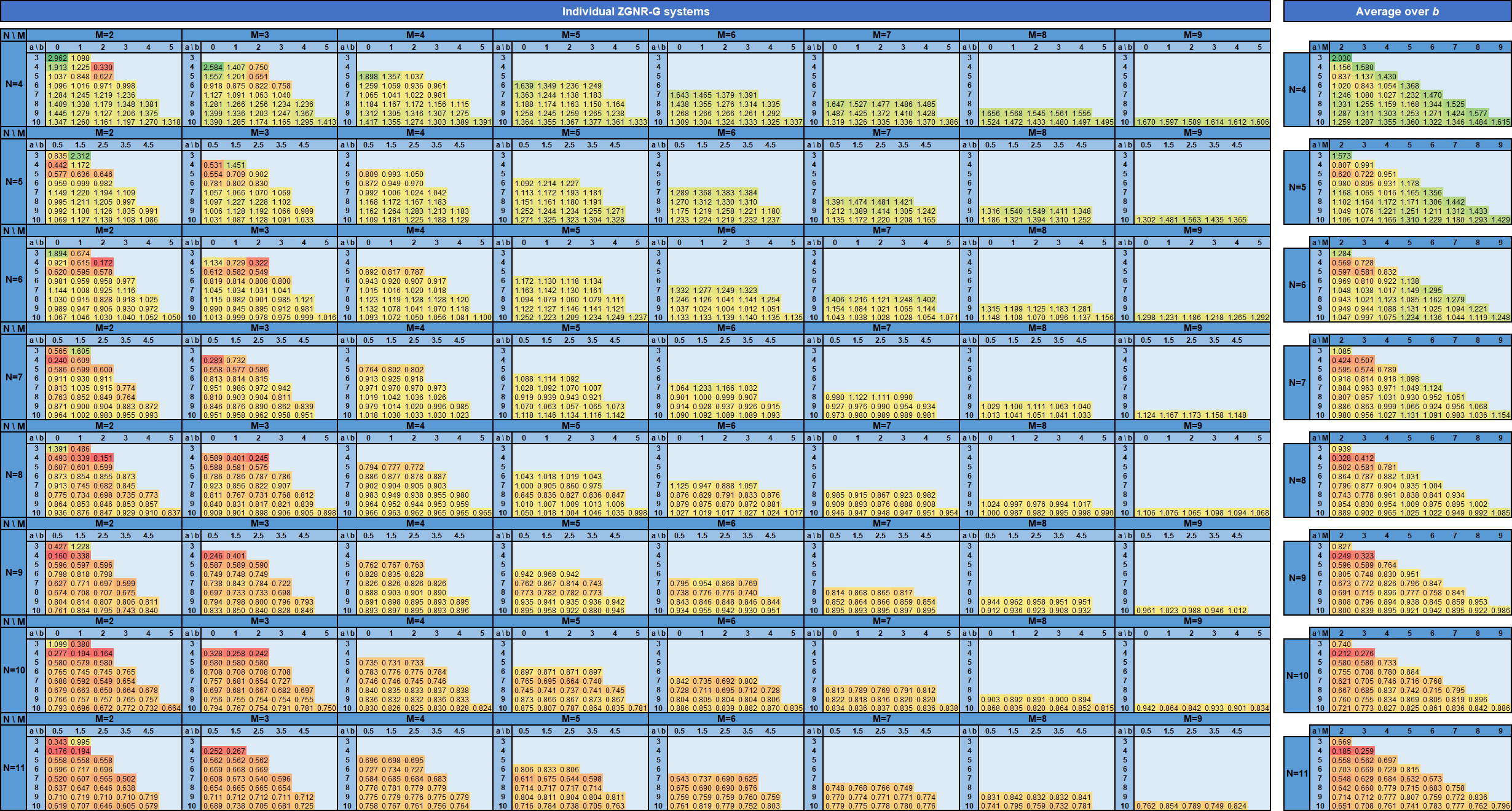}
    \caption{
    Dependency of the band gap (in units of eV) in the AFM state on the structural parameters in ZGNR-Gs, obtained from TB+U calculations.}
    \label{fig_SI_table_gap_size}
\end{sidewaysfigure}

\begin{sidewaysfigure}[ht!]
    \centering
    \includegraphics[width=\textwidth]{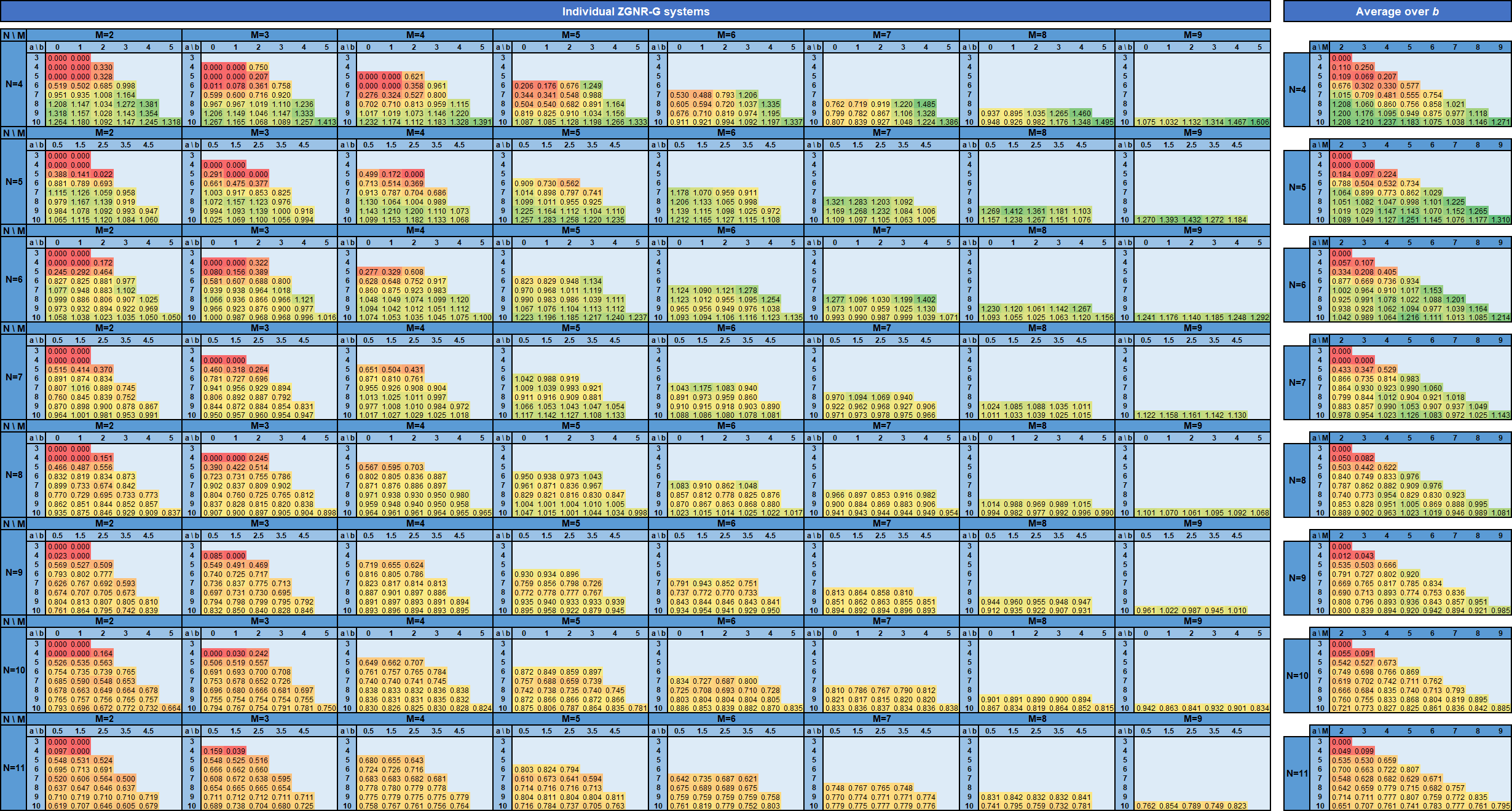}
    \caption{
    Dependency of the band gap opening (in units of eV) in the AFM state relative to the diamagnetic state on the structural parameters in ZGNR-Gs, obtained from TB+U calculations.}
    \label{fig_SI_table_gap_opening}
\end{sidewaysfigure}

\begin{sidewaysfigure}[ht!]
    \centering
    \includegraphics[width=\textwidth]{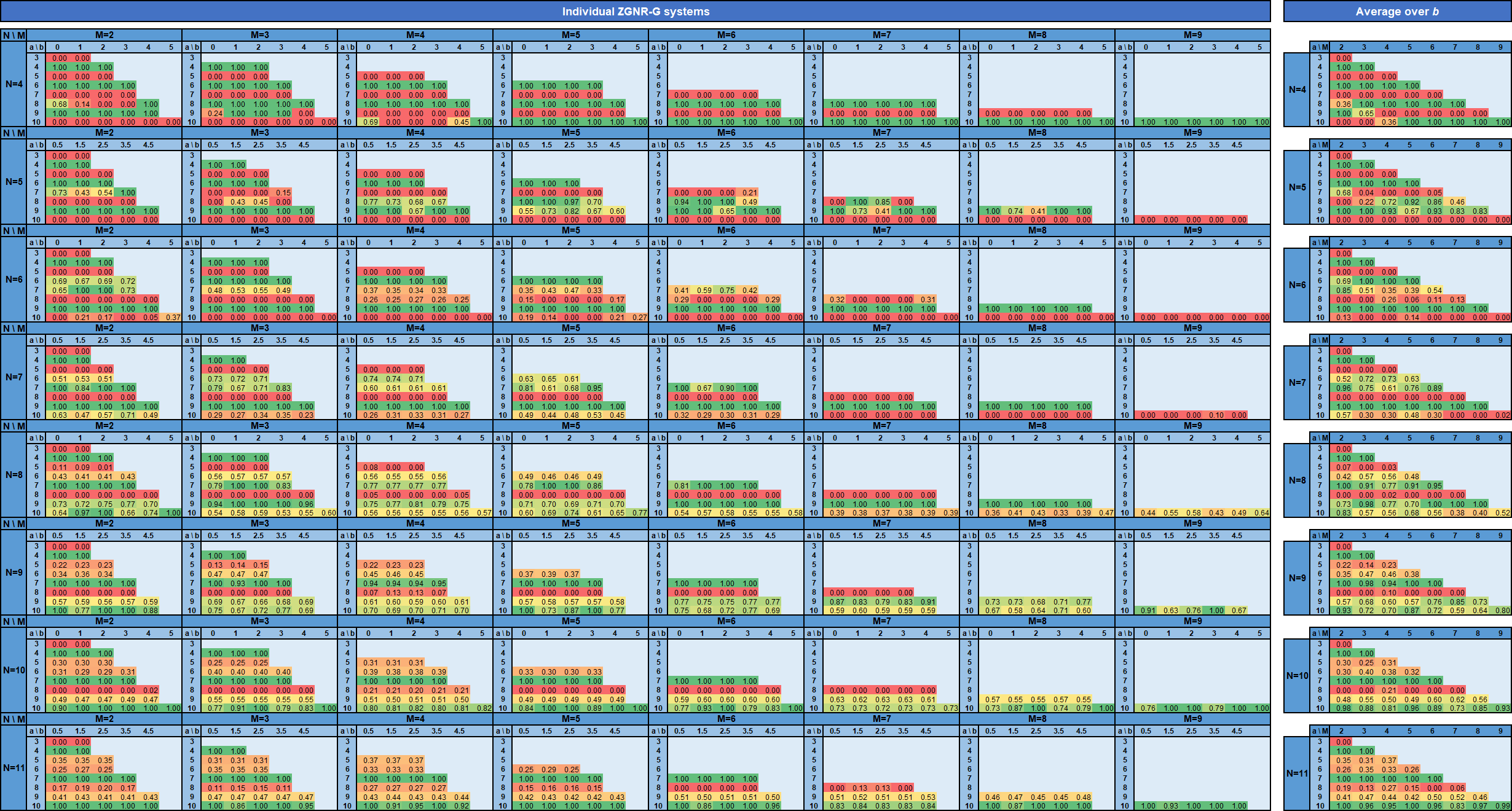}
    \caption{
    Dependency of the position of the band gap in the AFM state on the structural parameters in ZGNR-Gs, obtained from TB+U calculations. A value of 0 (1) corresponds to the $\Gamma$ ($X$) point at $k=0$ ($k=\pi$).}
    \label{fig_SI_table_gap_position}
\end{sidewaysfigure}

\clearpage
\subsection{Spin moments of ZGNR-Gs}
\label{sec_SI_magnetism}
\begin{figure*}[ht!]
    \centering
    \includegraphics[width=\textwidth]{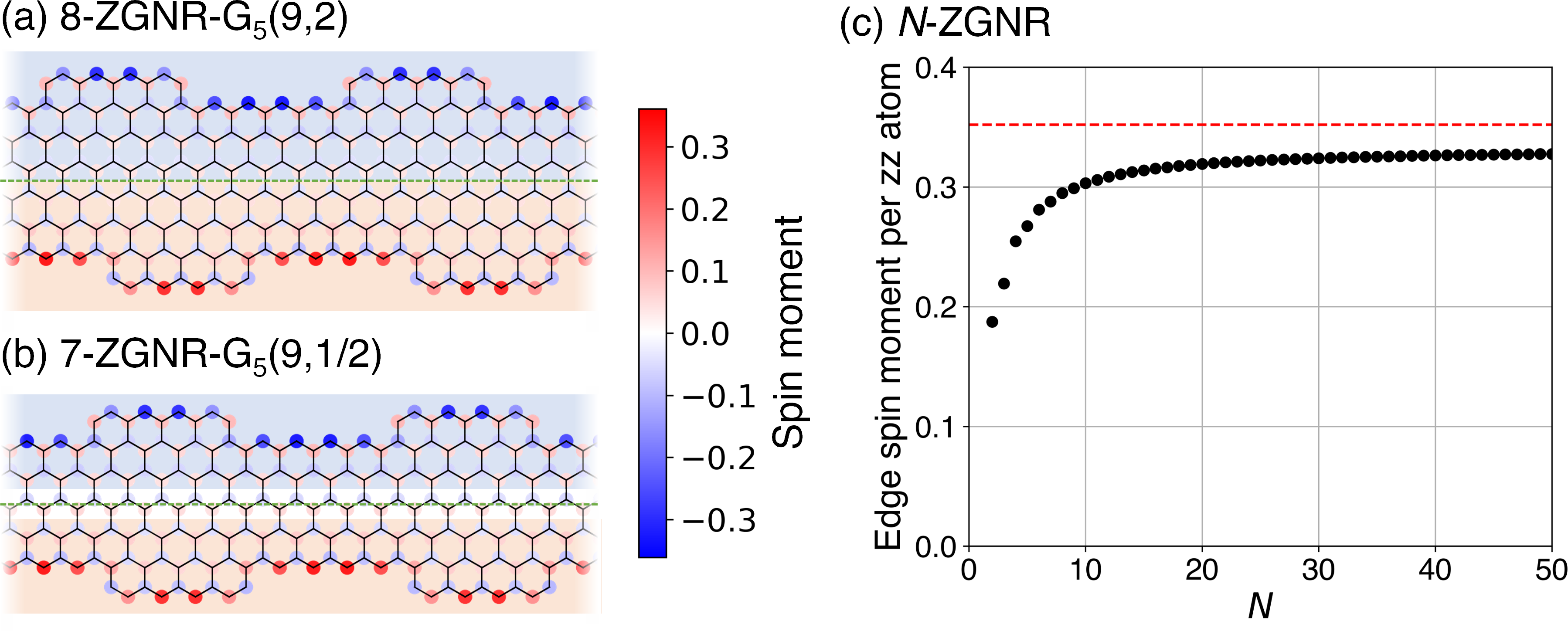}
    \caption{Edge spin moments. Schematic representation of how the total spin moment per zigzag edge is obtained in the case of (a) even $N$ and (b) odd $N$. The dashed green line corresponds to the center of the ribbon in the non-periodic direction; the shaded areas indicate the atoms that are summed up to obtain the total spin moment of the respective zigzag edge. (c) The absolute value of the spin moment per zigzag edge for ZGNRs as function of their width $N$ is normalized by the number of zigzag edge atoms per zigzag edge. The dashed red line corresponds to the maximum absolute spin moment in ZGNR.}
    \label{fig_SI_spin_per_edge}
\end{figure*}

\begin{sidewaysfigure}[ht!]
    \centering
    \includegraphics[width=\textwidth]{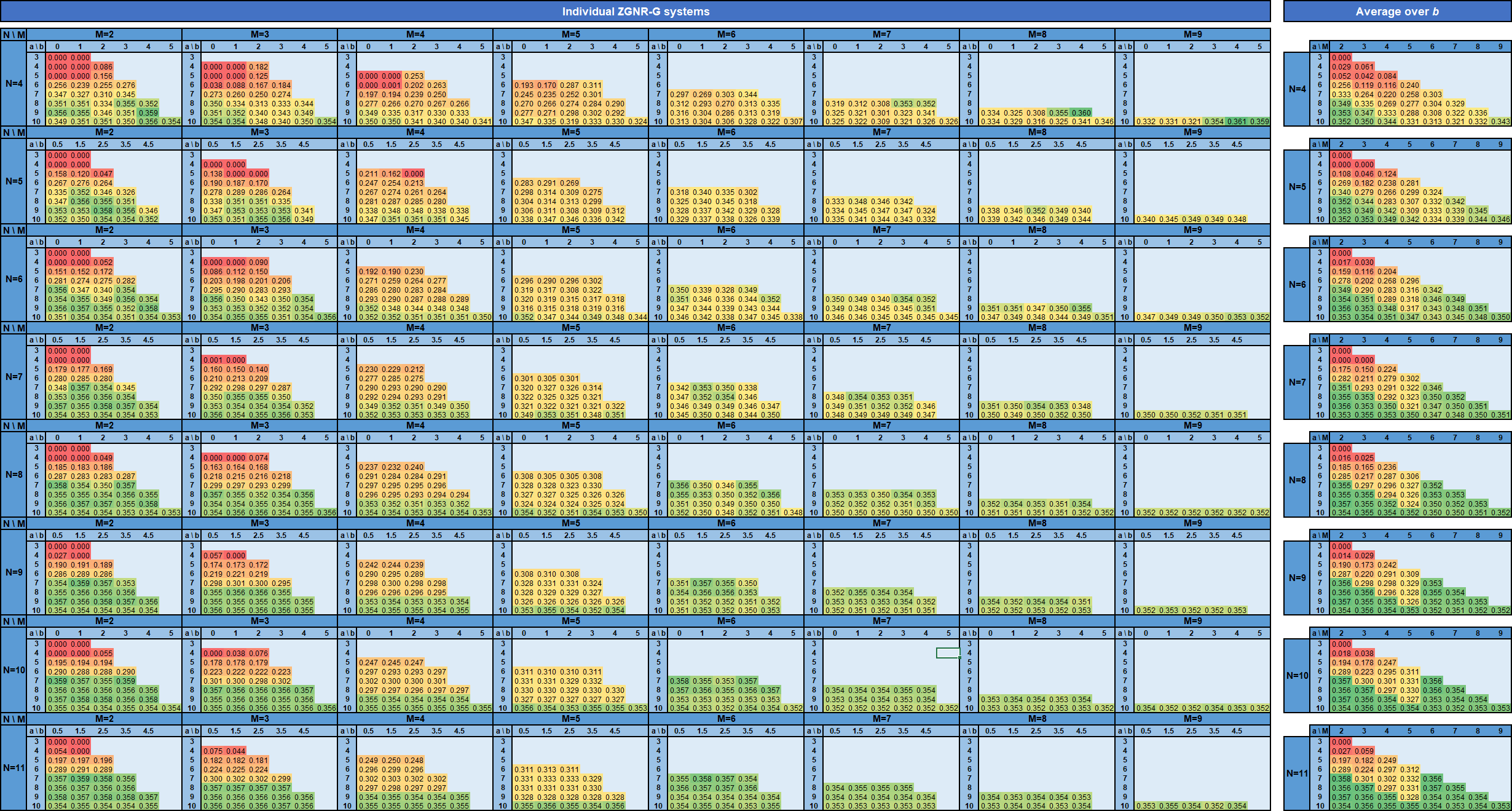}
    \caption{
    Dependency of the maximum absolute spin moment on the structural parameters in ZGNR-Gs, obtained from TB+U calculations.}
    \label{fig_SI_table_spin_max}
\end{sidewaysfigure}

\begin{sidewaysfigure}[ht!]
    \centering
    \includegraphics[width=\textwidth]{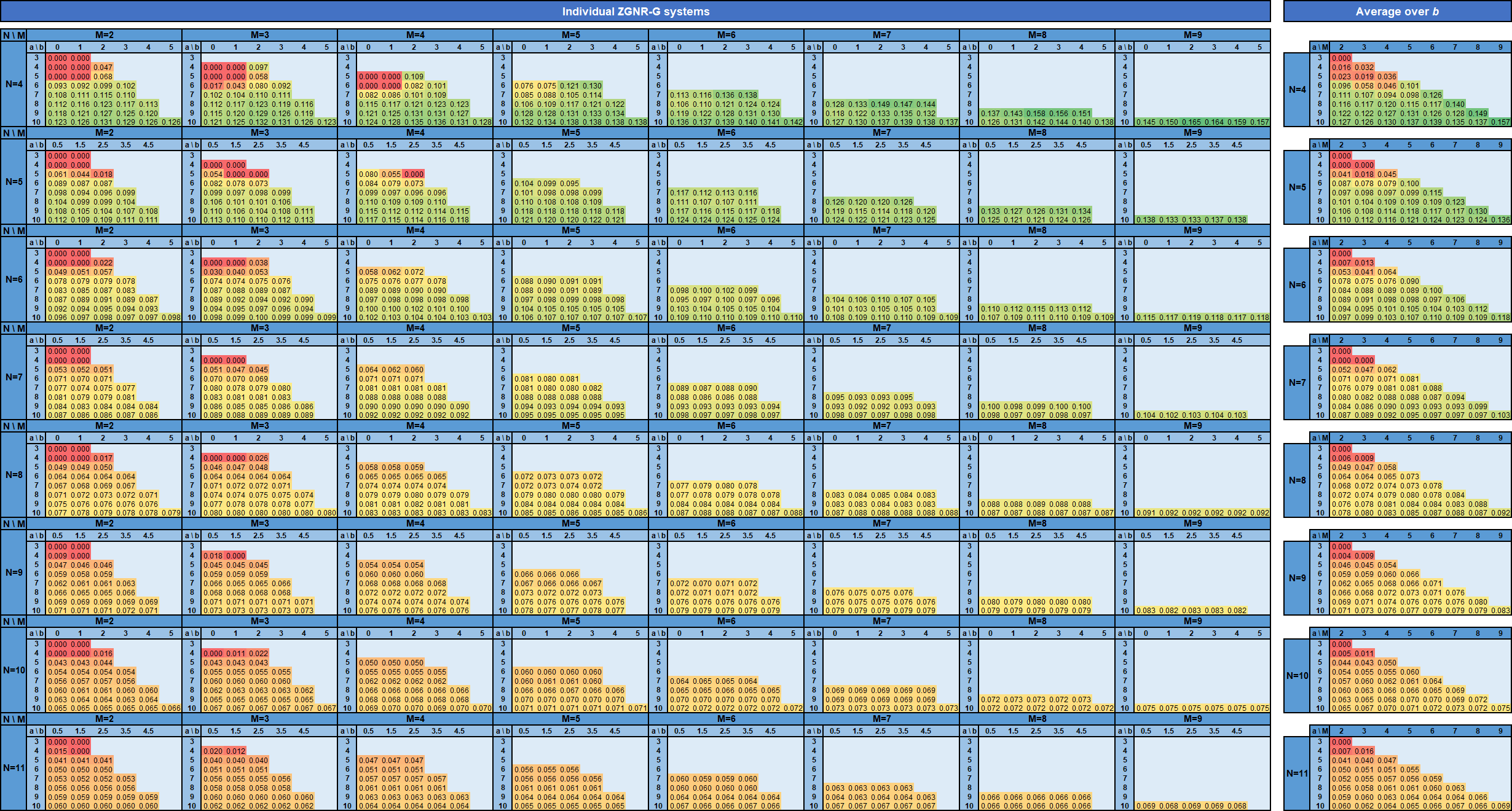}
    \caption{
    Dependency of the average absolute spin moment on the structural parameters in ZGNR-Gs, obtained from TB+U calculations.}
    \label{fig_SI_table_spin_average}
\end{sidewaysfigure}

\begin{sidewaysfigure}[ht!]
    \centering
    \includegraphics[width=\textwidth]{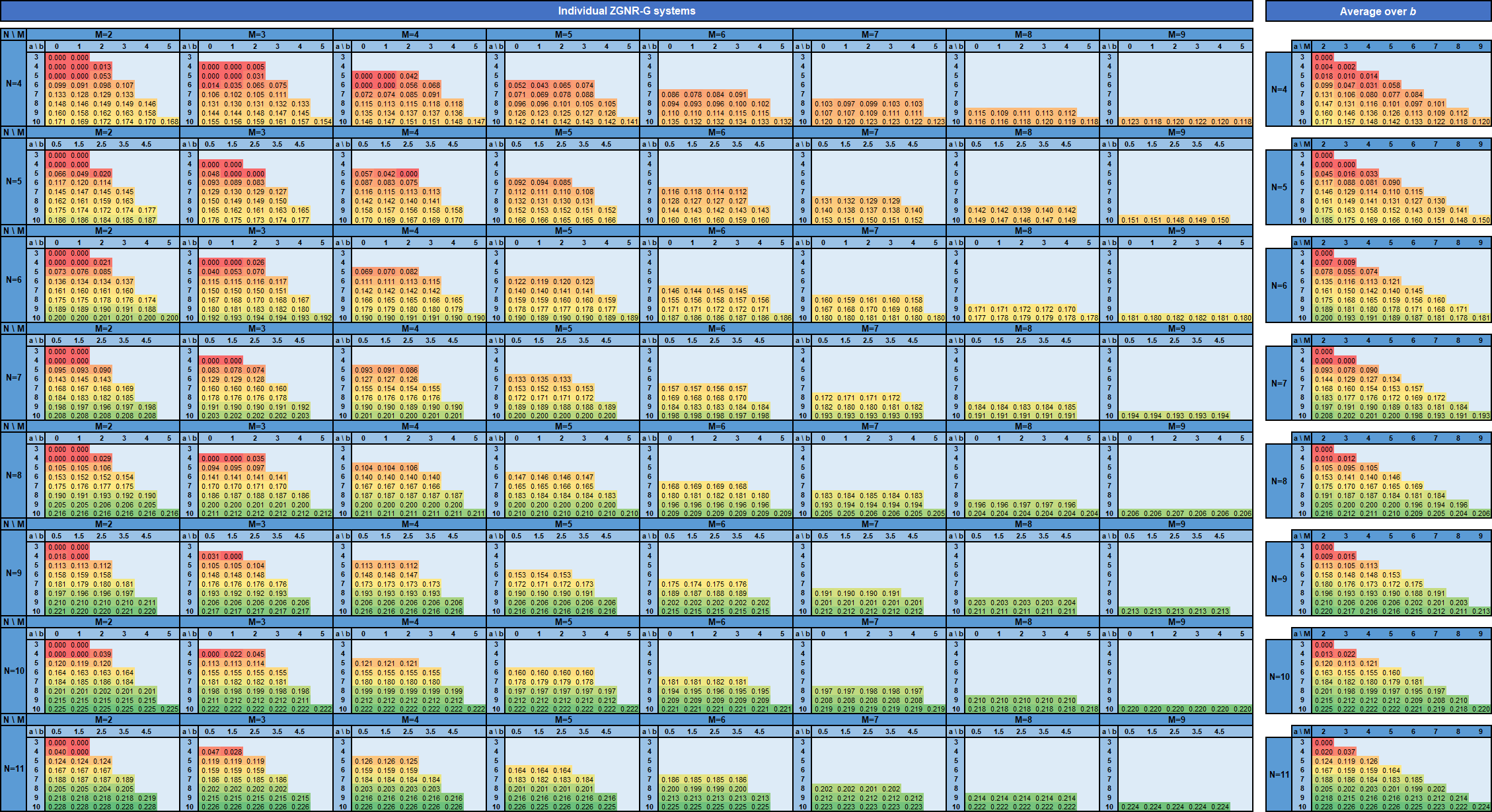}
    \caption{
    Dependency of the absolute total spin moment per zigzag edge of ZGNR-G on the structural parameters in ZGNR-Gs, obtained from TB+U calculations. The values are normalized by the number of zigzag edge atoms per zigzag edge.}
    \label{fig_SI_table_spin_per_edge}
\end{sidewaysfigure}

We analyzed the maximum absolute spin moment (see Fig.~\ref{fig_SI_table_spin_max}), the average absolute spin moment (see Fig.~\ref{fig_SI_table_spin_average}), and the absolute total spin moment per zigzag edge (see Fig.~\ref{fig_SI_table_spin_per_edge}) of ZGNR-Gs with AFM ordering on TB+U level of theory.
While the first two quantities can be understood intuitively, the third needs additional explanation.
As shown in Fig.~\ref{fig_SI_spin_per_edge}(a,b), the structure is split into an upper and lower part, assigning half of the atoms to the respective zigzag edge.
This way, the main contribution originates from the zigzag edge, while the contributions from atoms in the center cancel out due to the opposite spin moment in the sublattices.
This is easily possible for even $N$, as the two halves are cut between two zigzag rows.
However, the cut is through the central zigzag row for odd $N$, preventing the components from canceling out.
To counteract this, we neglect the central zigzag row when calculating the total spin moment per zigzag edge in odd $N$.
This approach successfully avoids the appearance of artifacts, as shown in Fig.~\ref{fig_SI_spin_per_edge}(c), where we observe a continuous change with $N$ for ZGNRs.
For large $N$, the value converges to $\approx 0.33$ per zigzag edge atom.
In ZGNR-Gs, the largest observed values are $\approx 0.23$ per zigzag edge atom.

\clearpage
\subsection{Inner vs. outer zigzag edge in the AFM state}
\begin{figure*}[ht!]
    \centering
    \includegraphics[width=0.5\textwidth]{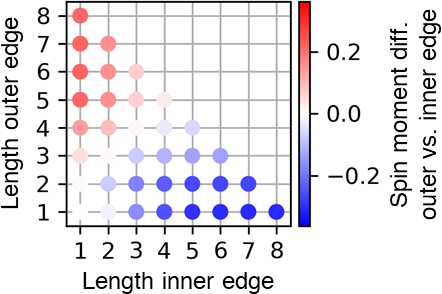}
    \caption{
    Difference between the maximum spin moment at the outer and inner zigzag edge. A negative (positive) value corresponds to the inner (outer) edge showing a larger spin moment. The crossover for equal spin polarization at both edges appears when the outer zigzag edge is one atom longer than the inner zigzag edge. The shown data is obtained by varying $a$, $M$, and $b$, averaged over all $N$ in the data set.}
    \label{fig_SI_outer_vs_inner_edge}
\end{figure*}

\begin{figure*}[ht!]
    \centering
    \includegraphics[width=0.5\textwidth]{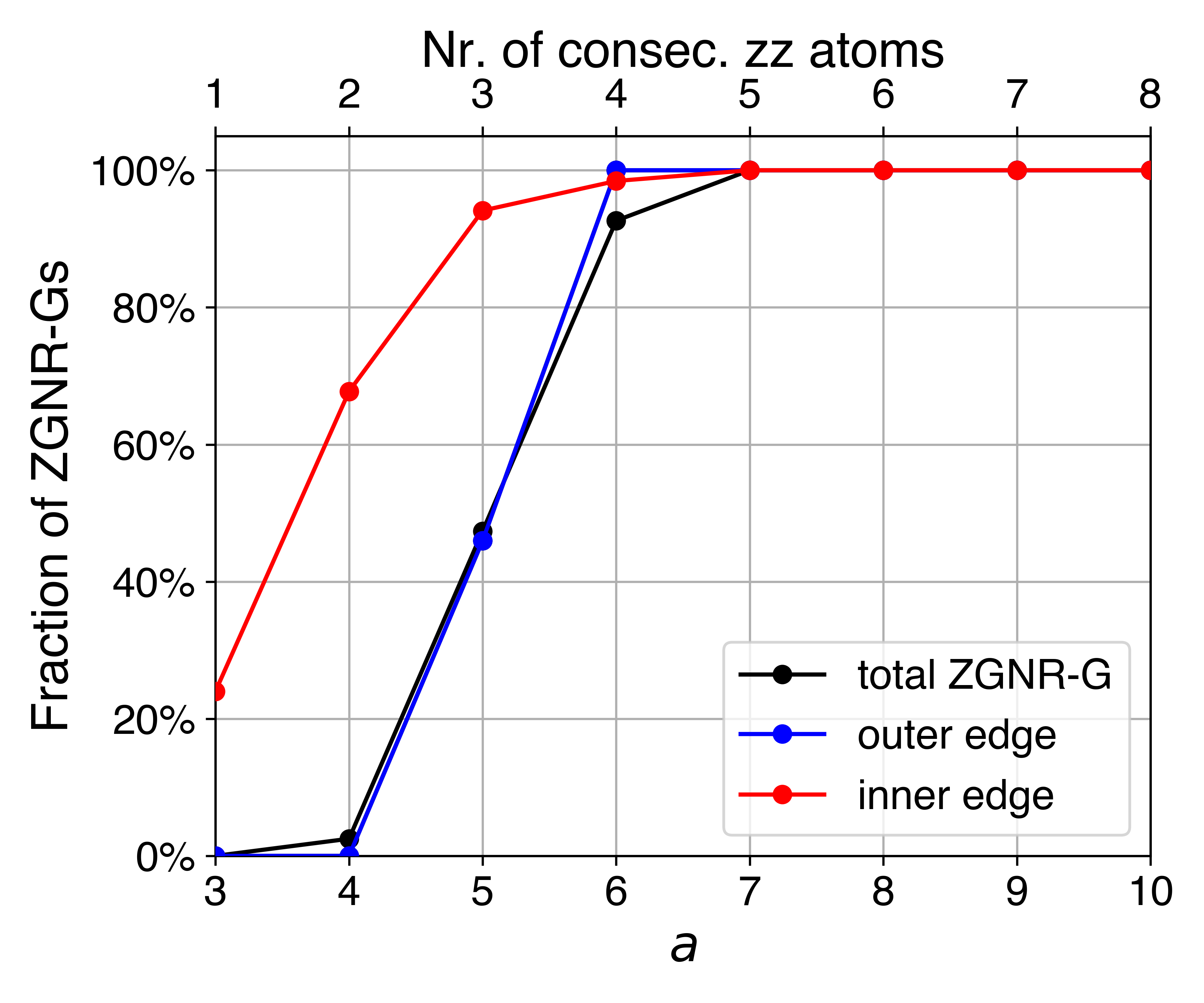}
    \caption{Dependency of the spin polarization on structural parameters. The fraction of the ZGNR systems with a strong AFM state at the DFT/HSE06 level if given as function of (primary x-axis) the parameter $a$ and (secondary x-axis) the number of consecutive zigzag edge atoms on the outer and inner edge. We define the AFM state as strong once the maximum spin moment reaches at least 50\% of the maximum spin moment in ZGNR systems.}
    \label{fig_SI_AFM_statistics_DFT}
\end{figure*}

\clearpage
\subsection{Topological properties of ZGNR-Gs}
% Figure: tables of Z2 values, depending on N, n, a, b and inversion center
\begin{figure*}[htp!] % pagewidth figure
    \centering
    \includegraphics[width=\textwidth]{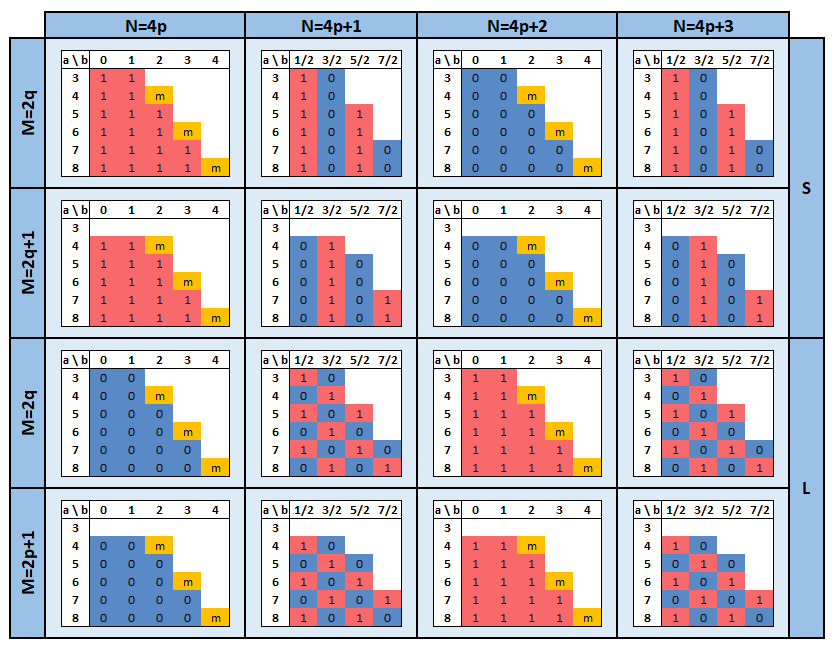}
     \caption{Dependency of topological properties on the structural parameters. The $\mathbb{Z}_2$ invariant for ZGNR-Gs is given, sorted by (outer columns) width $N$, (outer rows) gulf size $M$, (inner columns) gulf distance $a$, and (inner rows) gulf offset $b$ with integer $p = 1,2,\ldots$ and $q = 1,2,\ldots$. Topological insulators ($\mathbb{Z}_2=1$) are marked in red, trivial semiconductors ($\mathbb{Z}_2=0$) in blue, and metallic ribbons, as obtained in spin-restricted TB calculations, are indicated as "m" (yellow). The influence of the gulf size $M$ is demonstrated using $M=2$ and $M=3$ as examples for $M=2q$ and $M=2q+1$, respectively.}
    \label{fig_SI_Z2_table_full}
\end{figure*}

\newpage
\section*{References}
\providecommand{\latin}[1]{#1}
\makeatletter
\providecommand{\doi}
  {\begingroup\let\do\@makeother\dospecials
  \catcode`\{=1 \catcode`\}=2 \doi@aux}
\providecommand{\doi@aux}[1]{\endgroup\texttt{#1}}
\makeatother
\providecommand*\mcitethebibliography{\thebibliography}
\csname @ifundefined\endcsname{endmcitethebibliography}
  {\let\endmcitethebibliography\endthebibliography}{}

\end{document}